\documentclass[12pt,notitlepage]{article}

\usepackage{mathtools}
\usepackage{amsmath}         
\usepackage{amssymb}   
\usepackage{empheq}
\newcommand*\widefbox[1]{\fbox{\hspace{2em}#1\hspace{2em}}}
\usepackage{amsfonts}        
\usepackage{graphicx}        
\usepackage{authblk}
\usepackage{bbold} 

\usepackage{color}         
\usepackage{slashed}       
\usepackage{framed}        
\usepackage[]{empheq} 
\usepackage{cite}          
\usepackage{booktabs}      
\usepackage{multirow}    
\usepackage{youngtab}	    
\usepackage{pifont}        
\usepackage{bbm}           
\usepackage{tocloft}		   
\usepackage{setspace}	   
\usepackage{listings}       
\usepackage{comment}
\usepackage{soul}
\setstcolor{blue}

\usepackage{tikz}
\usepackage{tkz-euclide}
\usetikzlibrary{decorations.pathmorphing}	
\usetikzlibrary{shapes}	

\usetikzlibrary{shapes.geometric, positioning, fit}

\tikzset{every picture/.style={line width=1}}

\definecolor{c1}{rgb}{0.121569, 0.466667, 0.705882}
\definecolor{c2}{rgb}{1., 0.498039, 0.054902}
\definecolor{c3}{rgb}{0.172549, 0.627451, 0.172549}
\definecolor{c4}{rgb}{0.839216, 0.152941, 0.156863}
\definecolor{c5}{rgb}{0.580392, 0.403922, 0.741176}
\definecolor{c6}{rgb}{0.54902, 0.337255, 0.294118}
\definecolor{c7}{rgb}{0.890196, 0.466667, 0.760784}
\definecolor{c8}{rgb}{0.498039, 0.498039, 0.498039}
\definecolor{c9}{rgb}{0.737255, 0.741176, 0.133333}
\definecolor{c10}{rgb}{0.0901961, 0.745098, 0.811765}

\definecolor{KNgreen}{HTML}{1DB100}
\definecolor{KNorange}{HTML}{FEAE00}
\definecolor{KNblue}{HTML}{56C1FF}
\definecolor{KNgray}{HTML}{D5D5D5}
\definecolor{KNdkgray}{HTML}{5E5E5E}
\definecolor{KNred}{HTML}{EE220C}

\usepackage[margin=2cm]{geometry}   
\graphicspath{{figures/}}	        
\renewcommand{\tilde}{\widetilde}   

\newcommand{\ket}[1]{\left|#1\right\rangle}    

\let\oldenumerate\enumerate
\renewcommand{\enumerate}{
  \oldenumerate
  \setlength{\itemsep}{1pt}
  \setlength{\parskip}{0pt}
  \setlength{\parsep}{0pt}
}

\let\olditemize\itemize
\renewcommand{\itemize}{
  \olditemize
  \setlength{\itemsep}{1pt}
  \setlength{\parskip}{0pt}
  \setlength{\parsep}{0pt}
}

\usepackage[
	colorlinks=true,
	citecolor=black,
	linkcolor=black,
	urlcolor=blue,
	hypertexnames=false]{hyperref}

\renewcommand{\eqref}[1]{Eq.~(\ref{#1})}

\newcommand\Tstrut{\rule{0pt}{2.5ex}}         
\newcommand\Bstrut{\rule[-1.3ex]{0pt}{0pt}}

\newcommand{\nablaLR}{\overset{\leftrightarrow}{ \nabla}}
\newcommand{\hatn}{{\vec n}}

\newcommand{\vperp}{\vec{v}_\text{el}^\perp}
\newcommand{\vin}{\vec{v}_\text{in}^\perp}
\newcommand{\vzero}{\vec{v}_{0}^\perp}

\newcommand{\fperp}{\vec{f}_{1\rightarrow 2}^\perp}
\newcommand{\fpell}{{f}_{1\rightarrow 2}^{\parallel}}
\newcommand{\Aperp}{\vec{A}^\perp}
\newcommand{\Apell}{A^{\parallel}}

\newcommand{\On}[1]{\langle\mathcal{O}_{#1}\rangle}

\title{Simplified Spin Dependence in Dark Matter Direct Detection}

\vspace{.3cm}

\author[1]{Pierce Giffin\thanks{pgiffin@ucsc.edu}}
\affil[1]{\small\it Department of Physics and Santa Cruz Institute for Particle Physics, University of California Santa Cruz, Santa Cruz, CA 95064, USA}

\author[2,3]{Benjamin Lillard\thanks{blillard@psu.edu}}
\affil[2]{\small\it Institute for Gravitation and the Cosmos, The Pennsylvania State University, University Park, PA 16802, USA}
\affil[3]{\small\it Institute for Fundamental Science and Department of Physics, University of Oregon, Eugene, OR 97403, USA}

\author[1]{Pankaj Munbodh\thanks{pmunbodh@ucsc.edu}}

\author[3]{Tien-Tien Yu\thanks{tientien@uoregon.edu}}

\begin{document}
\maketitle

\begin{abstract} 
\normalsize 
The interactions of dark matter with Standard Model particles can be systematically studied in the language of effective field theories.  We investigate dark matter interactions with Standard Model particles, including spin-dependent interactions, for direct detection experiments and demonstrate that, although the scattering rate generally depends on multiple types of material response functions, certain linear combinations of these material response functions vanish if the initial and final electronic states share the same Hamiltonian. We also find that several other response functions vanish in parity-symmetric materials, making these systems as simple as isotropic detectors in some respects. Finally, we present the scattering rate for an anisotropic, possibly chiral detector, for generic dark matter-electron spin interactions. These relations reduce the number of independent response functions needed, thereby simplifying the computational complexity for a broad class of dark matter models. Our results provide a complete and efficient toolkit for analyzing electron recoil signals in diverse detector materials. 
\end{abstract}

\newpage
\tableofcontents

\section{Introduction}

Direct detection experiments provide an opportunity to study the microscopic interactions between dark matter (DM) and Standard Model (SM) particles by directly probing the local DM population in our galaxy. By measuring energy deposits from rare scattering events, these experiments provide powerful constraints on a wide range of DM models that are independent of the DM production mechanism and cosmological distribution. 
In this paper, we combine the complete nonrelativistic effective theory for DM-SM interactions with a fully three-dimensional model of the detector medium, so as to facilitate \textit{ab initio} calculations of the generic spin-dependent DM scattering rate for the next generations of anisotropic direct detection experiments. 

Traditionally, direct detection experiments have primarily focused on DM interactions with nuclei in the target material. This approach is particularly well-suited for probing Weakly Interacting Massive Particles (WIMPs),
and over the past four decades, experiments have achieved remarkable sensitivities~\cite{Akerib:2022ort, PICO:2019vsc, DarkSide-50:2022qzh, LZ:2024zvo, XENON:2025vwd}. However, despite these advances, no unambiguous DM signals have been observed. In response, DM–electron scattering has emerged as a promising strategy to explore alternative DM candidates, especially those with masses below that of the nucleon (sub-GeV DM)~\cite{Essig:2011nj,Essig:2022dfa}. 

The DM-electron scattering rate depends on the underlying DM model as well as the structure and SM properties of the detector medium. A broadly model-independent framework to describe this rate is provided by nonrelativistic 
effective field theory (EFT), which classifies interactions through Lagrangian operators that are invariant under rotations and Galilean velocity boosts. 
Previous studies have cataloged complete sets of invariant operators for DM-electron~\cite{Catena:2019gfa,Catena:2022fnk,Catena:2024rym,Liang:2024lkk,Liang:2024ecw} interactions, 
building on similar efforts for DM-nucleus interactions~\cite{Fan:2010gt,Fitzpatrick:2012ix,DelNobile:2018dfg}. In this EFT formalism, the Wilson coefficients of the operators provide a complete accounting of the relevant interactions between nonrelativistic DM and the detector. 
This formalism is sufficiently precise for DM in the galactic halo, where relativistic corrections on the order of $(v/c)^2 \sim 10^{-6}$ can be safely neglected. 

A key ingredient in computing the DM-electron scattering rate is a set of material-dependent response functions (or form factors), which encode the likelihood that momentum, energy, and possibly spin transferred from DM to SM particles will excite the target system into a final observable state. Importantly, these response functions depend solely on the SM dynamics, namely the initial and final state wavefunctions of the electrons, and can therefore be calculated independently of the dark sector.

Initial works on DM-electron scattering focused on spin-independent scattering in an isotropic material, which consists of a single EFT operator and depends on a single, scalar response function ~\cite{Essig:2011nj,Essig:2015cda,Derenzo:2016fse,Essig:2017kqs}.
However, this limits the applicability of the results to a subset of possible DM models, excluding possibilities such as spin-dependent interactions or DM models with a magnetic or anapole moment (see e.g.~Refs~\cite{Essig:2019kfe,Trickle:2019ovy,Trickle:2020oki,Liu:2021avx,Wang:2021oha,Esposito:2022bnu,Berlin:2025uka,Krnjaic:2025noj,Marocco:2025eqw,Gori:2025jzu,Hochberg:2025rjs} for some discussions of the direct detection of spin-dependent sub-GeV DM). 
Refs.~\cite{Catena:2019gfa,Catena:2021qsr} generalized the calculation of DM–electron scattering to a broader set of EFT 
operators and identified four additional response functions, under the simplifying assumption that the detector materials are isotropic, i.e., spherically symmetric in their response.
Refs.~\cite{Liang:2024lkk,Liang:2024ecw} expanded on this work to find that, with a redefinition of the material response functions, the squared matrix element can be organized into five terms that are a product of the DM response functions, which depend only on the Wilson coefficients of the EFT operators, and a linear combination of the material response functions. For spherically symmetric atomic targets, two of these response functions vanish.
In addition, they connect the nonrelativistic operators to the relativistic framework of low-energy EFT. 

Realistic materials, including those used or proposed for direct detection experiments, can exhibit anisotropy, and in these systems the response functions of Refs.~\cite{Catena:2019gfa,Liang:2024lkk,Liang:2024ecw} are insufficient to predict the DM scattering rate.
Anisotropy is a desirable detector property: if the DM scattering rate depends on the detector orientation, then an experiment can use the expected sidereal daily modulation to distinguish a possible DM signal from the various SM backgrounds~\cite{Hochberg:2016ntt, Hochberg:2017wce, Coskuner:2019odd, Geilhufe:2019ndy, Griffin:2020lgd, Blanco:2021hlm, Hochberg:2021ymx, Blanco:2022pkt, Boyd:2022tcn, Romao:2023zqf, Lillard:2023qlx, Lillard:2023cyy, Stratman:2024sng, Marocco:2025eqw, Griffin:2025wew, Abbamonte:2025guf}.
In this work, we extend the nonrelativistic formalism to accommodate spin-dependent interactions in \textit{ab initio} calculations for anisotropic detector targets. 
We find that, in addition to the response functions associated with isotropic detectors, the scattering rate receives a contribution from two additional vectorial response functions, consistent with Ref.~\cite{Catena:2021qsr}, raising the number of response functions to seven. 
Following the organizing principles in Ref.~\cite{Liang:2024ecw,Liang:2024lkk}, 
we expand on previous work to present for the first time the generic nonrelativistic DM-electron scattering rate in a generically anisotropic detector. 
 
Beyond generalizing the formalism, we identify a significant simplification for a broad class of materials. In systems where the initial and final state wavefunctions of the SM target particle are eigenstates of the same Hamiltonian, such as in atomic systems or single-particle excitations in crystals, we derive new relationships among the response functions. Specifically, we show that certain spin-dependent contributions can be expressed in terms of the scalar (spin-independent) form factor, up to boundary terms that typically vanish. These results reduce the number of independent response functions required to compute the scattering rate, thereby easing the computational burden for a wide range of DM interaction models and detector materials.

Exceptions to this simplification are found in processes like non-adiabatic Migdal scattering in molecules~\cite{Blanco:2022pkt} and crystals~\cite{Esposito:2025iry}, which are driven by corrections to the Born-Oppenheimer approximation. Here, the target’s initial and final states are governed by different Hamiltonians due to dynamic coupling between nuclear and electronic degrees of freedom. Collective excitations, such as phonons or plasmons, can also violate the assumptions underpinning our simplification (see {\it e.g.}~\cite{Catena:2024rym}). Nevertheless, for many cases of practical interest, especially atomic DM–electron scattering and single-particle excitations in molecules and crystals, the condition that the initial and final states are governed by the same Hamiltonian, $\hat{H}_i = \hat{H}_f$, provides a good approximation.

We begin in Section~\ref{sec:formalism} by extending previous derivations of the DM–electron scattering rate to anisotropic media, identifying the full set of response functions that contribute and clarifying how they appear in the rate. 
We conclude this section with a modestly simplified version of the scattering amplitude, for generic DM--SM spin couplings, in a completely anisotropic detector. 
The expression includes five linear combinations of SM response functions, rather than the seven that might have been expected from~\cite{Catena:2021qsr}. 
Next, Section~\ref{sec:simplified} demonstrates that two of these combinations vanish in parity-symmetric materials, even if the DM-SM particle interaction violates parity. 
Finally, in Section~\ref{sec:adiabatic}, we derive a novel relationship between the spin-dependent and spin-independent form factors, in the case where the initial and final electronic states are eigenstates of the same Hamiltonians. In the multiparticle systems where this is a good approximation, another linear combination of response functions vanishes. 
In the simplest cases, where all three conditions are satisfied, we show that even the most generic DM-SM spin interactions depend only on the ``spin-independent'' form factor. 
Section~\ref{sec:casestudies} provides a few concrete examples of DM models, with a spin-0 or spin-$\frac{1}{2}$ DM particle coupled to electrons via spin-0 and/or spin-1 mediators. We conclude with an inelastic scalar DM model, as a simple example where all of the DM response functions are nonzero.

\section{Formalism for DM-SM Scattering}
\label{sec:formalism}
In this section, we review the formalism for calculating the DM scattering rate as well as the definitions of the material response functions. Section~\ref{sec:responsefunctions} concludes with our expression for generically spin-dependent DM--SM interactions in an anisotropic detector medium with unpolarized spins.

\subsection{Scattering Rate and Matrix Element}
The total DM--SM scattering rate in a direct detection experiment is given by 
\begin{align}
R_{1 \rightarrow 2}  &= \frac{N_\text{cell} \rho_\chi /m_\chi}{64 \pi^2 m_\chi^2 m_T^2} \int\! d^3 q \, d^3 v\, f_\chi(\vec v) \, \delta\!\left( E+ \frac{q^2}{2 m_\chi} - \vec q \cdot \vec v \right) \left| \mathcal M_{1 \rightarrow 2} (\vec q, \vec v ) \right|^2 ,
\label{eq:rate}
\end{align}
where $N_\text{cell}$ is the number of target cells/molecules/etc.~in the detector; $m_\chi$ and $\rho_\chi$ are the DM particle mass and local mass density; $f_\chi(\vec v)$ is the DM velocity distribution; $E =  E_2 - E_1$ and $\vec q$ are the energy and momentum transferred from DM to the SM target particle, which has mass $m_T$;  and $\mathcal M_{1 \rightarrow 2}$ is the transition amplitude 
\begin{align}
\mathcal M_{1 \rightarrow 2} ( \vec q , \vec v) &= \int\! \frac{ d^3 k }{(2 \pi)^3} \tilde \psi_2^\star(\vec k + \vec q) \, \mathcal M_{\rm free}( \vec q, \vec v, \vec k)\, \tilde \psi_1(\vec k) .
\label{def:M12}
\end{align}
Here $\psi_1$ and $\psi_2$ are the initial and final states of the scattered SM particle, 
with $\tilde{\psi_i}(\vec k) = \langle \vec k | \psi_i \rangle$ the momentum space wavefunctions, 
and $ \mathcal M_{\rm free}( \vec q, \vec v, \vec k)$ is the amplitude for free particle scattering~\cite{Essig:2015cda}.
To transform from momentum space to position space, we use the normalization convention
\begin{align}
\psi(\vec x) &= \int\! \frac{d^3 k}{(2\pi)^3} e^{i \vec k \cdot \vec x} \tilde \psi(\vec k)\, ,
&
\tilde \psi(\vec k) &= \int\! d^3 x\, e^{-i \vec k \cdot \vec x}\psi(\vec x)\, .
\label{def:fourier}
\end{align}

In the non-relativistic limit, Galilean boost invariance dictates that $\mathcal M_{\rm free}(\vec q, \vec v, \vec k)$ should depend only on $\vec q$, the momentum transfer, and 
the elastic transverse relative velocity $\vperp$, defined as
\begin{align}
\label{eq:vperp}
\vperp &\equiv \vec v - \frac{ \vec q}{2 \mu_{\chi T} } - \frac{\vec k }{m_T} \, ,
\end{align}
where $\mu_{\chi T}$ is the reduced mass of the DM-SM system.
Following the notation of~\cite{Liang:2024lkk}, we can expand $\mathcal M_{\rm free}$  to linear order in $\vperp$, 
\begin{align}
\mathcal M_{\rm free}( \vec q, \vec v, \vec k) &= \mathcal M_{\rm free}(\vec q, \vperp) \simeq  \mathcal M_{\rm free}(\vec q, 0) + \vperp \cdot \left( \nabla_{\vperp} \mathcal M_{\rm free}(\vec q, \vperp) \Big|_{\vperp = 0} \right) 
\\&\equiv  \mathcal M_S(\vec q) + \vperp \cdot \vec{\mathcal M}_V(\vec q)\, ,
\end{align}
where $\mathcal M_S$ and $\vec{\mathcal M}_V \equiv \nabla_{\vperp} \mathcal M(\vec q, \vperp) \big|_{\vperp = 0}$ each depend only on $\vec q$. 
In this expression, $\vperp$ contains all of the $\vec k$ dependence in the free particle amplitude $\mathcal M_{\rm free}$. Note that since $\mathcal{M}_S$ and $\mathcal{M}_V$ are independent of the SM particle momentum $\vec k$, they can be pulled out of the integral in \eqref{def:M12} and 
\begin{equation}
    \mathcal M_{1 \rightarrow 2} ( \vec q , \vec v) = \mathcal M_S(\vec q)f_S(\vec q) + \vec{\mathcal M}_V(\vec q)\cdot\vec{f}_V(\vec q)\, .
\end{equation}
All of the $\vec k$-dependent terms are encoded in a scalar form factor, $f_S$, and a vector form factor, $\vec f_V$, 
\begin{align}
f_S(\vec q) &\equiv \int\! \frac{d^3 k}{(2\pi)^3} \tilde \psi_2^\star(\vec k + \vec q) \cdot 1 \cdot   \tilde \psi_1(\vec k)  = \int\! d^3x\, e^{i \vec q \cdot \vec x} \psi_2^\star(\vec x) \cdot 1 \cdot \psi_1(\vec x) , 
\\
\vec f_V(\vec q, \vec v) &\equiv \int\! \frac{d^3 k}{(2\pi)^3} \tilde \psi_2^\star(\vec k + \vec q) \cdot \vperp(\vec k) \cdot   \tilde \psi_1(\vec k)  = \int\! d^3x\, e^{i \vec q \cdot \vec x} \psi_2^\star(\vec x) \cdot \left( \vzero + \frac{i\nabla}{m_T}  \right)  \cdot \psi_1(\vec x) , 
\end{align}
where for convenience we define
\begin{align}
\vzero &\equiv \vec v - \frac{\vec q}{2 \mu_{\chi T} }\, . 
\end{align}
The detector response functions, which we define shortly, depend on bilinear products of $(f_S, \vec f_V)$ and their complex conjugates.

Specializing to the familiar case of spin-independent SM--DM interactions, the $\mathcal M_V$ term vanishes from the rate, leaving
\begin{align}
R_{1 \rightarrow 2}  &= \frac{N_\text{cell} \rho_\chi /m_\chi}{64 \pi^2 m_\chi^2 m_T^2} \int\! d^3 q \, d^3 v\, f_\chi(\vec v) \, \delta\!\left( E+ \frac{q^2}{2 m_\chi} - \vec q \cdot \vec v \right) \left| \mathcal M_{S} (\vec q, \vec v ) \right|^2 |f_S(\vec q)|^2 ,
\label{eq:SIrate}
\end{align}
reproducing the known results~\cite{Essig:2015cda}. 
For an anisotropic detector, the scattering rate $R_{1 \rightarrow 2}$ is a function of the detector orientation, which can be described by a rotation operator $\mathcal R \in SO(3)$ acting on some initial configuration of the detector:
\begin{align}
R_{1 \rightarrow 2}(\mathcal R)  &= \frac{N_\text{cell} \rho_\chi }{64 \pi^2 m_\chi^3 m_T^2} \int\! d^3 q \, d^3 v\, f_\chi(\vec v) \, \delta\!\left( E+ \frac{q^2}{2 m_\chi} - \vec q \cdot \vec v \right) \mathcal R \cdot \left| \mathcal M_{1 \rightarrow 2} (\vec q, \vec v ) \right|^2 .
\label{eq:rateR}
\end{align}
Equivalently, the scattering rate can be evaluated in the frame of a fixed detector, with the inverse rotation operator acting on the 3d DM velocity distribution, 
\begin{align}
R_{1 \rightarrow 2}(\mathcal R)  &= \frac{N_\text{cell} \rho_\chi }{64 \pi^2 m_\chi^3 m_T^2} \int\! d^3 q \, d^3 v\, \left( \mathcal R^{-1} \cdot f_\chi(\vec v) \right) \, \delta\!\left( E+ \frac{q^2}{2 m_\chi} - \vec q \cdot \vec v \right)   \left| \mathcal M_{1 \rightarrow 2} (\vec q, \vec v ) \right|^2 .
\label{eq:rateRinv}
\end{align}
In both cases, $\mathcal R \cdot f$ is an active rotation applied to the function $f(\vec u)$, i.e.
\begin{align}
\mathcal R \cdot f(\vec u) &\equiv f(\mathcal R^{-1} \vec u). 
\end{align}

The amplitude $\mathcal M$ can be approximated by its non-relativistic limit with negligible loss of accuracy, as long as the SM target particle is also non-relativistic.
In DM--nuclear scattering, the lab-frame speed of the nucleus is typically negligible, 
but  electronic bound states typically have the fastest speeds in the DM--detector system: in terms of the fine structure constant ($\alpha < 0.01$), $|\vec k|/m_e \simeq |\vec v_e| \sim \alpha\,c$.
However, even these $\mathcal O(v^2) \sim \alpha^2 \sim 10^{-4}$ relativistic corrections for DM--electron scattering are negligible, especially compared to the astrophysical uncertainties on the local DM velocity distribution.
For the remainder of the discussion, we will focus on electronic targets but the general framework holds for other targets as well.

\subsection{Expansion in Response Functions} \label{sec:responsefunctions}

Now we turn our attention to $\mathcal M_S$ and $\vec{ \mathcal M}_V$, which encode how the DM--SM scattering depends on both the DM model parameters and the SM physics of the detector material. 
If the detector (or the DM) has polarized spins, then the free particle amplitude may include dot products and cross products between $\vec f_V$, $\vec q$, and the spin polarization vectors. 
In a typical direct detection experiment, however, the spins of the electrons and nuclei are unpolarized, as is the DM. 
Similarly, the usual detector designs might measure the energy transfer $E = E_2 - E_1$, but not the spin state of the scattered SM particle. 
With this setup, the measured event rate depends only on the spin averaged and summed squared amplitude, 
\begin{align}
\overline{|\mathcal M_{1 \rightarrow 2} |^2} &\equiv \frac{1}{2 j_\chi + 1} \frac{1}{2 j_\text{SM} + 1} \sum_{\text{spins}(i)} \sum_{\text{spins}(f) } |\mathcal M_{1 \rightarrow 2} |^2,
\end{align}
where $j_\chi$ is the DM particle spin (0, $\frac{1}{2}$, 1 for scalar, fermion, or vector boson DM), and $j_\text{SM}$ is the spin of the SM target particle. 

The spin-summed $\overline{|\mathcal M_{1 \rightarrow 2} |^2}$ depends on the scalar products of $\vec q$, $\vec f_V$ and $\vec f_V^\star$, 
\begin{align}
\overline{|\mathcal M_{1 \rightarrow 2} |^2} \supset |f_S|^2,~~ |\vec f_V|^2, ~~| \vec q \cdot \vec f_V|^2, ~~ \vec{q} \cdot (\vec f_V \times \vec f_V^\star), ~~
f_S (\vec q \cdot \vec f_V^\star)\, .
\end{align}
For scenarios in which the SM and DM spins are unpolarized and unmeasured, there are no other vectors in the system. 
Note that any of these terms may receive subleading corrections with additional scalar powers of $ q^2$, $\vec q \cdot \vzero$, or $|\vzero|^2$; 
however, because $\vec f_V$ is already linear in $\vperp$, 
bilinear combinations such as $\vzero \cdot \vec f_V$ are already higher order in the $v \ll c$ expansion.
Following this logic, Ref.~\cite{Liang:2024lkk} finds that $\overline{|\mathcal M_{1 \rightarrow 2}|^2}$ can be expanded into five terms: 
\begin{align}
\overline{|\mathcal M_{1 \rightarrow 2} |^2} &= a_0 | f_S|^2 + a_1|\vec f_V|^2 
+ \frac{a_2}{q^2/m_e^2} \left| \frac{\vec q}{m_e} \cdot \vec f_V \right|^2
+ i a_3 \frac{\vec q}{m_e} \cdot (\vec f_V \times \vec f_V^\star)
+ 2 \,\text{Im}\left[ a_4 f_S \left(\frac{\vec q}{m_e}  \cdot \vec f_V^\star \right) \right] .
\label{eq:amp2Liang}
\end{align}
The DM response functions, $a_{0,\ldots, 4}(q^2)$, encode all of the relevant microphysical details of the DM particle model through the Wilson coefficients of the non-relativistic operators (see Table~\ref{tab:NRops}). We work through a few specific examples in Section~\ref{sec:casestudies}, for fermionic and bosonic DM models.
Like $f_S$, the $a_{0,\ldots, 4}$ depend on the momentum transfer $\vec q$, and $\vec f_V$ is the only term that also depends on the DM velocity $\vec v$.

The exact forms of the $a_i$ depend on the spin of the DM, but they share some general features.
The first four $a_i$ ($a_{0,1,2,3}$) are explicitly real-valued, but $a_4$ can be complex, in the case of inelastic DM scattering (i.e.~the DM initial and final states are different particles, $m_\chi \neq m_\chi'$). 
Both $a_3$ and $a_4$ are generated entirely by interferences between pairs of Wilson coefficients. $a_{0,1,2}$, on the other hand, include terms proportional to $|c_i|^2$ as well as $c_i^\star c_j$.

\renewcommand{\arraystretch}{2}
\begin{table}
    \centering
    \begin{tabular}{|l|l|l|} 
    \hline 
        \phantom{$_1$}$\mathcal{O}_1 = \mathbb{1}_\chi \mathbb{1}_e$ &
          \phantom{$_1$}$\mathcal{O}_3 = \mathbb{1}_\chi \left (\frac{i\vec{q}}{m_e}\times\vec{v}^\perp_{\rm el}\right )\cdot \vec{S}_e$ &
          \phantom{$_1$}$\mathcal{O}_4 = \vec{S}_\chi\cdot \vec{S}_e$ \\
    \hline
          \phantom{$_1$}$\mathcal{O}_5 = \vec{S}_\chi\cdot \left (\frac{i\vec{q}}{m_e}\times\vec{v}_{\rm el}^\perp \right ) \mathbb{1}_e$ &
          \phantom{$_1$}$\mathcal{O}_6 = \left (\vec{S}_\chi\cdot  \frac{\vec{q}}{m_e} \right ) \left ( \frac{\vec{q}}{m_e}\cdot\vec{S}_e\right )$ &
          \phantom{$_1$}$\mathcal{O}_7 = \mathbb{1}_\chi \vec{v}^\perp_{\rm el}\cdot\vec{S}_e$ \\
    \hline
          \phantom{$_1$}$\mathcal{O}_8 = \vec{S}_\chi\cdot\vec{v}^\perp_{\rm el} \mathbb{1}_e$ &
          \phantom{$_1$}$\mathcal{O}_9 = -\vec{S}_\chi\cdot \left (\frac{i\vec{q}}{m_e}\times\vec{S}_e\right )$ &
          $\mathcal{O}_{10} = \mathbb{1}_\chi \frac{i\vec{q}}{m_e}\cdot \vec{S}_e $ \\
    \hline
          $\mathcal{O}_{11} = \vec{S}_\chi\cdot \frac{i\vec{q}}{m_e} \mathbb{1}_e$ &
          $\mathcal{O}_{12} = -\vec{S}_\chi\cdot (\vec{v}^\perp_{\rm el}\times \vec{S}_e)$ &
          $\mathcal{O}_{13} = \left (\vec{S}_\chi\cdot\vec{v}^\perp_{\rm el} \right )\left (\frac{i\vec{q}}{m_e} \cdot \vec{S}_e \right )$\\
    \hline
          $\mathcal{O}_{14} = \left (\vec{S}_\chi\cdot\frac{i\vec{q}}{m_e}\right ) \left (\vec{v}^\perp_{\rm el}\cdot\vec{S}_e \right ) $ &
          $\mathcal{O}_{15} = \vec{S}_\chi\cdot\frac{\vec{q}}{m_e} \left [ \frac{\vec{q}}{m_e}\cdot (\vec{v}^\perp_{\rm el}\times\vec{S}_e) \right ]$ &
          {}\\
    \hline 
    \end{tabular}
    \caption{List of effective non-relativistic operators for spin-0 and spin-1/2 DM coupling to an electron following the conventions of Refs.~\cite{Fan:2010gt,Fitzpatrick:2012ix}. Here we identify the DM and electron spin operators as $\vec{S}_\chi$ and $\vec{S}_e$, respectively along with the identity operators $\mathbb{1}_\chi$ and $\mathbb{1}_e$.}
    \label{tab:NRops}
\end{table}
\renewcommand{\arraystretch}{1}

An alternative way of defining the vector material response function is given by~\cite{Catena:2021qsr,Liang:2024lkk} 
\begin{align}
\vec f_{1 \rightarrow 2}(\vec q) &\equiv \int\! \frac{d^3 k}{(2\pi)^3} \tilde \psi_2^\star(\vec k + \vec q) \cdot \frac{+i \vec k}{m_T} \cdot   \tilde \psi_1(\vec k)  = \int\! d^3x\, e^{i \vec q \cdot \vec x} \psi_2^\star(\vec x) \cdot \left(  \frac{-i\nabla}{m_T}  \right)  \cdot \psi_1(\vec x)\, , 
\label{eq:Fconvert}
\end{align}
so that $\vzero$ contains all of the dependence on the DM velocity $\vec v$. Note
$
\vec f_V(\vec q, \vec v) = \vzero \, f_S(\vec q) - \vec f_{1 \rightarrow 2}(\vec q).
$
From $f_S(\vec q)$ and $\vec f_{1 \rightarrow 2}(\vec q)$, 
the scattering rate can be expressed as a linear combination of the DM response functions $a_{0,\ldots, 4}$ and $\vec v$-independent material response functions $B_i(\vec q)$~\cite{Catena:2021qsr}: 
\begin{align}
B_1 &= |f_S(\vec q)|^2 , 
&
B_2 &= \frac{\vec q}{m_e} \cdot  \vec A(\vec q), \nonumber
\\
B_3 &= | \vec f_{1 \rightarrow 2}(\vec q) |^2 , 
& 
B_4 &= \left| \frac{ \vec q}{m_e} \cdot \vec f_{1 \rightarrow 2}(\vec q) \right|^2 , 
\end{align}
\begin{align}
B_5 &= i \frac{ \vec q}{m_e} \cdot \left( \vec f_{1 \rightarrow 2}(\vec q) \times \vec f_{1 \rightarrow 2}^\star(\vec q) \right)=\frac{2 \vec q}{m_e} \cdot \left[ \text{Re}(\vec f_{1\to 2}) \times \text{Im}(\vec f_{1\to 2}) \right] \, , 
\label{def:B5}
\end{align}
where $\vec A(\vec q) \equiv f_S(\vec q) \vec f_{1 \rightarrow 2}^\star(\vec q)$.
Note that $B_1$, $B_2$ and $B_4$ are not independent: 
\begin{align}
|B_2|^2 &= \left(  \frac{\vec q}{m_e} \cdot  \vec A^\star \right) \left(  \frac{\vec q}{m_e} \cdot  \vec A \right) 
= B_1 \left(  \frac{\vec q}{m_e} \cdot  \vec f_{1 \rightarrow 2}^\star \right) \left(  \frac{\vec q}{m_e} \cdot  \vec f_{1 \rightarrow 2} \right) 
= B_1 B_4,
\label{eq:B124}
\end{align}
but the phase of $B_2$ cannot be extracted from $B_1$ and $B_4$.

For anisotropic detectors, we also need to include the vector response functions: 
\begin{align}
\vec B_6 &= \vec A(\vec q), 
&
\vec B_7 &= \frac{\vec q}{m_e} \times \vec A(\vec q).
\end{align}
Thus, the most general form of the squared amplitude is given by
\begin{empheq}[box=\widefbox]{align}
\overline{|\mathcal M_{1 \rightarrow 2} |^2} &= a_0 B_1 
+ a_1 \left[ |\vzero|^2 B_1 + \frac{m_e}{\mu_{\chi e} } \text{Re}\,  B_2 + B_3 - 2 \,\text{Re}\left( \vec v \cdot \vec B_6\right) \right]
\nonumber\\ &~~
+ a_2 \frac{m_e^2}{q^2} \left[ \left( \frac{\vec q \cdot \vzero}{m_e} \right)^2 B_1 - 2 \left( \frac{\vec q \cdot \vzero}{m_e} \right) \text{Re}\,  B_2 + B_4 \right] 
\nonumber\\ &~~
+ a_3 \left[ B_5 - 2 \, \text{Im}\left( \vec v \cdot \vec B_7 \right) \right]
- 2 \, \text{Re}\left( a_4 \right) \text{Im}\,B_2\nonumber\\ &~~
+2\,\text{Im}(a_4)\left[\left( \frac{\vec q \cdot \vzero}{m_e} \right) B_1 - \text{Re}\, B_2\right]\, . 
\label{eq:a01234B1234567}
\end{empheq}
For models of elastic DM, the last line is identically zero as $\text{Im}(a_4) = 0$.
Since $\vec q \cdot \vec B_7  = 0$, the vectorial response functions only appear as the combination $\vec v \cdot \vec B_{6,7}$ in the squared amplitude.\footnote{Dot and cross products of $\vec q$ with $\vec B_6$ are already defined as $B_{2}$ and $\vec B_7$, respectively, and $\vec q\cdot\vec B_7=\vec A\cdot(\vec q\times\vec q)=0$.}
Ref.~\cite{Liang:2024ecw} provides the values of $a_{0, \ldots, 4}(c_i)$ as functions of the Wilson coefficients $c_i$ in their Table~4.

\subsection{Connecting DM Microphysics to the Detector Response}

\begin{figure}
\centering
\includegraphics[width=\textwidth]{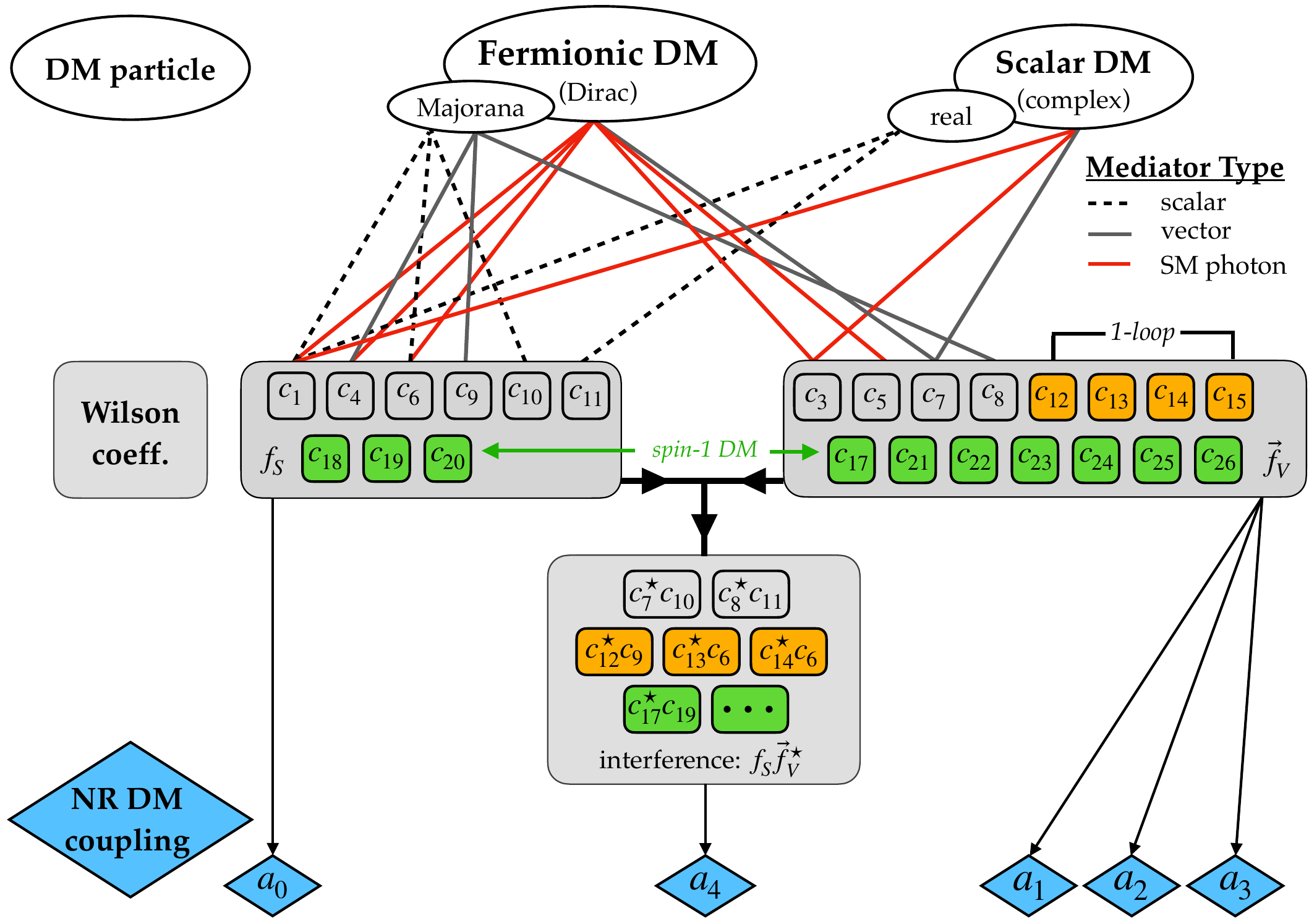}
\caption{
The mapping of example UV DM models ({\bf ovals}) to the $c_i$ Wilson coefficients, the 
material response functions $f_S$ and $\vec f_V$ ({\bf{\color{KNdkgray} gray rectangles}}), and 
the NR DM couplings $a_{0,\ldots,4}$ ({\bf{\color{KNblue}blue diamonds}}).  
The colored lines join the UV models (scalar real/complex DM or fermionic Dirac/Majorana DM) to the Wilson coefficients that they generate at tree level for dimension-4 operators, with the color of the line distinguishing between scalar ({\bf dashed black}) and vector ({\bf solid}) mediators. 
The SM photon ({\bf\color{KNred}red}) is a special case of vector mediator as it only couples to the electron through the vector current. 
Any $c_i$ that can be generated via the SM photon can also be generated by a generic BSM vector mediator.
All lines belonging to ``Majorana" (or ``real") also belong to ``Dirac" (or ``complex").
Note that the coefficients in {\bf\color{KNorange}orange} {\color{KNorange}$c_{12},\ldots,c_{15}$} can be generated at one-loop as shown in Figure \ref{fig:loops}.
Shown in {\bf{\color{KNgreen} green}} are the coefficients {\color{KNgreen}$c_{17},\ldots,c_{26}$} that can only be generated by a spin-1 DM particle (see Table 1 of \cite{Liang:2024ecw} for the operators they correspond to). 
The coefficients $c_i$ are classified by their contribution to the scalar form factor $f_S$ ({\bf left {\color{KNdkgray}gray rectangle}}) or to the vector form factor $\vec{f}_V$ ({\bf right {\color{KNdkgray}gray rectangle}}).
In the {\bf central {\color{KNdkgray}gray box}}, we show all the five possible interference terms for scalar and fermionic DM. We denote in {\bf\color{KNgreen}green} the interference terms that are generated by spin-1 DM. 
In the lower part of the diagram, we show the relation between the $c_i$ coefficients and the $a_{0,\ldots,4}$ coefficients.  }
\label{fig:tikz} 
\end{figure}

Eq.~(\ref{eq:a01234B1234567}) factorizes the SM physics of the detector, captured by the $B_{1 \ldots 7}(\vec q)$ response functions, from any dependence on the DM model parameters or the DM velocity, $\vec v$. 
Conveniently, the expansion in $\vec f_V$ from Ref.~\cite{Liang:2024lkk} keeps the DM model dependence contained within five coefficients $a_{0, \ldots, 4}$, rather than the seven we would naively expect from the number of $B_i(\vec q)$ functions.

Each $a_i$ depends on quadratic combinations of Wilson coefficients, corresponding to Lagrangian operators such as those listed in Table~\ref{tab:NRops}. Whether or not a given operator can appear in the Lagrangian depends on the spin of the DM particle and the spin of the off-shell mediator that couples the SM and DM sectors. In the simplest examples, when the mediator is a (spin-0) scalar, only $a_0$ is nonzero, and the only response function contributing to the scattering rate is $B_1$. If the mediator is a vector boson, then some of the spin-dependent interactions are permitted even if the DM itself is a scalar.
To keep track of which $a_i$ coefficients can be nonzero in each scenario, Figure~\ref{fig:tikz} depicts the mapping between DM particle models and the SM response functions, for spin-0 and spin-$\frac{1}{2}$ DM candidates. We also show which Wilson coefficients are generated by spin-1 DM candidates. 
Some features are immediately obvious: for example, the $a_4$ term is generated entirely by interference between pairs of Wilson coefficients, and it would vanish in any analysis that considers only one Wilson coefficient at a time. The spin-independent $a_0$ term only requires the scalar form factor $f_S$ while $a_{1,\ldots,4}$ require the contribution of the vector form factor $\vec f_V$.
Only renormalizable tree-level couplings between the mediator and the SM/DM particles are shown in the figure. Higher dimensional operators, like the SM anapole moment for Majorana DM, would generate additional connections in the graph.

\begin{figure}[htbp]
    \centering
    \begin{minipage}{0.2\textwidth}
    \includegraphics[width=\linewidth]{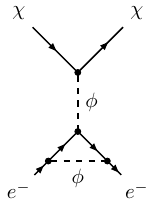}
  \end{minipage}
    \hfill
   \begin{minipage}{0.2\textwidth}
    \includegraphics[width=\linewidth]{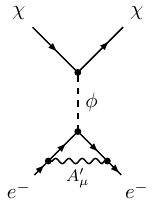}
  \end{minipage} 
    \hfill
    \begin{minipage}{0.2\textwidth}
    \includegraphics[width=\linewidth]{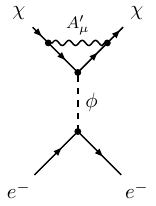}
  \end{minipage}
    \hfill
    \begin{minipage}{0.25\textwidth}
    \includegraphics[width=\linewidth]{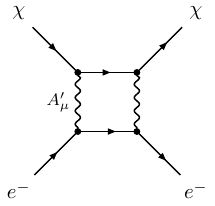}
  \end{minipage}

\caption{From left to right, the first two loop diagrams generate $\mathcal{O}_{15}$; the second and third diagrams generate $\mathcal{O}_{14}$ and $\mathcal{O}_{13}$,  respectively; 
and the last diagram generates $\mathcal{O}_{12}$. 
The $\mathcal O_{13,14}$ contributions from these loops vanish as $q \rightarrow 0$.
}
\label{fig:loops}
\end{figure}

Before moving on, we briefly comment on how the operators $\mathcal{O}_{12}$ to $\mathcal{O}_{15}$ may be generated at one-loop from the diagrams displayed in Fig.~\ref{fig:loops}. First, notice that all of these operators require the DM to be fermionic due to the appearance of the spin $\vec{S}_\chi$. From the form of the operators and the non-relativistic reduction of the spinor currents, it is straightforward to understand how they are generated. In the case of $\mathcal{O}_{13}, \mathcal{O}_{14}, \mathcal{O}_{15}$, the DM current and the electron current are not contracted with each other. A scalar mediator $\phi$ is needed since it does not carry any Lorentz index, whereas in the case of $\mathcal{O}_{12}$ a vector mediator is needed to couple the two currents. 
To obtain the non-relativistic $\vec{S}_{\chi,e}\cdot\vec{v}^\perp_{\rm el}$
current appearing in $\mathcal{O}_{14},\mathcal{O}_{13}$, the relativistic current $(\bar{u}_{\chi,e}\gamma^\mu\gamma^5 u_{\chi,e})(\bar{u}_{e,\chi}\gamma^5 u_{e,\chi})$ must contract with an external momentum. The external momentum $(p+p^\prime)_\mu$ (or $(k+k^\prime)_\mu$) is afforded by the fermion propagators in the loop, while the fermion current is afforded by the vector $A^\prime_\mu$ mediator. For $\mathcal{O}_{12}$, $ ( \bar{u}_\chi \gamma_\mu \gamma_5u_\chi)(\bar{u}_e\sigma^{\mu\nu}u_e)$ must contract with a non-zero momentum to maintain Lorentz invariance. The first two loop diagrams can generate $\mathcal{O}_{15}$. For the first diagram, $\mathcal{O}_{15}$ is generated from the contraction of $(\bar{u}_\chi\gamma^5u_\chi)(\bar{u}_e\sigma^{\mu\nu}u_e)$ or $(\bar{u}_\chi\gamma^5u_\chi)(\bar{u}_e\sigma^{\mu\nu}\gamma^5u_e)$ with two momenta. For the second diagram, $\mathcal{O}_{15}$ is generated from the contraction of $(\bar{u}_\chi\gamma^5u_\chi)(\bar{u}_e\gamma^\mu u_e)$ with an external momentum.

Rather than keep the original $\vec B_{i = 1\ldots 7}$ response functions of Ref.~\cite{Catena:2021qsr}, we find it more convenient to separate the vectorial form factor $\vec f_{1 \rightarrow 2}(\vec q)$ into a radial component, $\hat q\cdot \vec f_{1 \rightarrow 2}$, and a tangential component, $\vec f^\perp_{1 \rightarrow 2}$:
\begin{align}
\vec f_{1 \rightarrow 2} &\equiv  (\fpell)  \, \hat q +  \fperp , 
&
 \fpell  &\equiv \hat q \cdot \vec f_{1 \rightarrow 2}, \nonumber
\\
\vec A &\equiv (\Apell) \, \hat q + \vec A^\perp,
&
\Apell &  
= f_S (\fpell)^\star, 
\end{align}
satisfying $\fperp \cdot \vec q = 0$ and $\Aperp \cdot \vec q = 0$ everywhere.
Not only does this simplify the mapping between Wilson coefficients and response functions, but it also makes it manifest which of the response functions should vanish in isotropic systems.

With this notation, the material response functions $B_{i \geq 2}$ become:
\begin{align}
B_2 &= \frac{q}{m_e} \Apell,
&
B_3 &= | \fpell|^2 + |\fperp|^2 = 
B_3^\parallel + B_3^\perp, 
&
B_3^{(\perp, \parallel)} &\equiv |\vec f_{1 \rightarrow 2}^{\,(\perp, \parallel)}|^2 , \nonumber
\\
B_4 &= \frac{q^2}{m_e^2} B_3^\parallel , 
&
B_5 &= \frac{2 \vec q}{m_e} \cdot \left[ \text{Re}(\fperp) \times \text{Im}(\fperp) \right] , 
&
\vec B_7 &= \frac{\vec q }{m_e} \times \Aperp.
\label{eq:altB1234567}
\end{align}
Finally, the combination $\vec v \cdot \vec B_6$ that appears in the scattering rate simplifies to:
\begin{align}
\vec v \cdot \vec B_6 &= v_\text{min}(q) \Apell + \vec v \cdot \Aperp,
\end{align}
where 
\begin{align}
v_\text{min}(q) &\equiv \frac{E_2 - E_1}{q} + \frac{q}{2 m_\chi}. 
\end{align}

With the exception of the $a_4$ components, all of the terms in \eqref{eq:amp2Liang} that depend on $\fpell$ but not $\fperp$ appear in the following combination:
\begin{align}
B_H &\equiv \left( \hat q \cdot \vzero \right)^2 B_1 - 2 \left( \hat q \cdot \vzero \right) \text{Re}\,  \Apell + B_3^\parallel .
\end{align}
This combination of response functions has an important physical significance: we demonstrate in Section~\ref{sec:adiabatic} that if the initial and final state wavefunctions are eigenstates of the same Hamiltonian, then $B_H \rightarrow 0$ vanishes identically. 

In conclusion, noting that $|\vzero|^2 - (\hat q \cdot \vzero)^2 = v^2 - v_\text{min}^2$, 
the squared scattering amplitude for a generic detector material is:
\begin{align}
\overline{|\mathcal M_{1 \rightarrow 2} |^2} &= a_0 B_1 
+ a_1 \left[ \left( v^2 - v_\text{min}^2  \right) B_1     - 2 \,\text{Re}\left(\vec v \cdot \Aperp \right) 
 + B_3^\perp
 \right]
+ (a_1 + a_2)  B_H
\nonumber\\ &~~
+ a_3 \left[ B_5 - 2 \, \text{Im}\left( \vec v \cdot \vec B_7 \right) \right]
+ \frac{2 q}{m_e} \Big[ \text{Im}(a_4) \left[ \left(\hat q \cdot \vzero \right) B_1 - \text{Re}\,\Apell \right] 
-   \text{Re}\left( a_4 \right)   \text{Im}\,\Apell 
\Big] \, ,
\label{eq:aEtcB0}
\end{align}
with $\text{Im}(a_4) \rightarrow 0$ for elastic DM scattering, i.e.~when the incoming and outgoing DM states are the same. 
This is the same generic squared amplitude as \eqref{eq:a01234B1234567}, reorganized to make it easier to track which terms vanish in each simplifying regime studied in Section~\ref{sec:simplified}.

\section{Symmetries and Simplifying Regimes} \label{sec:simplified}
In all but the simplest detector models, the analytic forms of the wavefunctions $\psi_{1,2}(\vec x)$ are not known exactly. In semiconductors and molecules, and (once the electron-electron Coulomb potentials are taken fully into account) in atoms other than hydrogen, 
the energy eigenstates $\psi_{1,2}$ must generally be obtained numerically. 
For this reason, the form factors $f_S(\vec q)$ and $\vec f_{1 \rightarrow 2}(\vec q)$ often represent the most time-consuming part of the calculation: a numerically derived $\psi_{1,2}$ must be integrated over $d^3x$ (or momentum space) once for every value of $\vec q$. If $\psi_{1,2}$ are not spherically symmetric, a theory prediction may struggle under the weight of all these integrals.
In this light, it is unfortunate that, in addition to $B_1$, we must apparently also calculate several new spin-dependent response functions. 

Thankfully, many of the relevant, physical systems obey symmetries or properties that allow for simplifications. In this section, we show how parity, isotropy, and adiabaticity simplify the expressions for spin-dependent scattering, to something that is almost as simple as in the spin-independent case.
Section~\ref{sec:parity} demonstrates that the $a_3$ and $a_4$ terms in \eqref{eq:aEtcB0} vanish if the detector material is parity-symmetric, or invariant under reflections.
Section~\ref{sec:iso} investigates two types of isotropic systems: those that are isotropic because their constituent particles are spherically symmetric, as in atomic systems, and those that have become isotropic simply because the unit cells of the material are not aligned, i.e.~the target is a fluid or a glass.
Finally, in Section~\ref{sec:adiabatic} we consider systems where the initial and final state single particle wavefunctions $\psi_i$ and $\psi_f$ are eigenstates of the same Hamiltonian. Using their equations of motion, we derive a relationship between $\vec f_V$ and $f_S$, resulting in the $a_2$ and $a_4$ terms vanishing.

\subsection{Parity Symmetry} \label{sec:parity}
The spin-averaged squared amplitude depends on five linear combinations of seven material response functions. In parity-symmetric (nonchiral) materials, however, we demonstrate that several of these response functions vanish:
specifically, the $a_3$ and $a_4$ terms in \eqref{eq:aEtcB0}, and the related response functions $B_5$ and $\text{Im}(\vec B_7)$.
This simplification has previously been observed in atomic systems~\cite{Catena:2019gfa,Liang:2024ecw}, but attributed imprecisely to rotational invariance rather than parity.

Parity, or central inversion symmetry, 
is common in the symmetry groups of crystals, including those with anisotropic response functions~\cite{Blanco:2021hlm}. Thus, these results are relevant to a wide class of detector material candidates. 
The main counterexamples include chiral crystals~\cite{Arvanitaki:2021wjk}, or detector systems where a strong external $\vec B$ field breaks the parity symmetry.

\paragraph{Parity Eigenstates}
We begin our discussion by focusing on Hamiltonians that commute with the parity operation $P$,
\begin{align}
[P, H]  = 0\, .
\end{align}
Here $[P, H] \equiv PH - HP$ is the commutator, and the parity operator $P$ acts on the position space or momentum space coordinate systems to set $\vec x \rightarrow - \vec x$ and $\vec k \rightarrow - \vec k$, respectively.
Let $\psi_{1,2}$ be eigenstates of Hamiltonians $H_{1,2}$ of the form
\begin{align}
H_i = - \frac{\nabla^2}{2m_i} + V_i(\vec x) ,
\end{align}
i.e.~with no coupling to an external magnetic field.
At this stage, we do not need to assume $m_1 = m_2$ or $V_1 = V_2$, merely that $H_i$ is real. 
A parity-symmetric potential $V(\vec x)$ is invariant when reflected through a particular point: without loss of generality, we choose this point to be the origin in position space, so that $V(\vec x) = V(- \vec x)$. 

The eigenstates of $H_i$, $\ket{\psi_i}$, form an orthogonal basis of the Hilbert space. Any basis function can be redefined by multiplying it by a constant phase, $\ket{\psi_i} \rightarrow e^{i \delta} \ket{\psi_i}$; 
using this freedom, we can ensure that $\psi_i(\vec x_0)$ is real valued at some arbitrary point $\vec x_0$ where $\psi_i(\vec x_0) \neq 0$. 
Because each $H(\vec x)$ is a real-valued operator on position space functions $\psi(\vec x)$, 
$H \psi_i = E_i \psi_i$, this phase choice $e^{i \delta}$ ensures that for non-degenerate eigenstates
\begin{align}
\psi_i(\vec x) = \text{Re}\,\psi_i(\vec x)  
\label{eq:psireal}
\end{align}
for all $\vec x$.

This argument holds even if $H$ has degenerate eigenstates. Take for example the special case where $V(\vec x) = 0$ and $[H, k] = 0$. Momentum eigenstates $|\vec k\rangle \sim \exp(i \vec k \cdot \vec x)$ are not strictly real-valued. 
However, a generic final state with energy $E = k^2/2m$ is a linear combination of the degenerate energy eigenstates with fixed $|\vec k| = k$, including the pair of momentum states $|\pm\! \vec k\rangle$; therefore,
the Hilbert space can be spanned by real-valued eigenstates
\begin{align}
\psi_{\pm \vec k}(\vec x) \propto |\vec k \rangle \pm |\!-\! \vec k \rangle , 
&&
\psi_+(\vec x) \sim \cos(\vec k \cdot \vec x), 
&&
\psi_-(\vec x) \sim \sin(\vec k \cdot \vec x)\, .
\end{align}
Similar examples can be constructed for spherically symmetric potentials, where $H$ commutes with the generators of rotation. 

Once we have established a basis of real-valued energy eigenstates $\psi_i$ in position space, we can specialize to parity-symmetric Hamiltonians. 
If $[P, H] = 0$, a nondegenerate $\psi_i$ must also be a parity eigenstate: 
\begin{align}
P \psi_i = (\pm 1) \psi_i\, ,
\end{align}
i.e.~$\psi_i(-\vec x) = \pm \psi_i(\vec x)$.
If $\psi_i$ is not a $P$ eigenstate ($P \psi_i \neq \pm \psi_i$), but it still satisfies $H \psi_i = E_i \psi_i$ for a parity-symmetric $H$, then there must be a degenerate energy eigenstate $\psi_i'$ such that 
\begin{align}
P \psi_i = \psi_i'\, .
\end{align}
Otherwise, $[P, H] \psi_i \neq 0$. Given these two degenerate states $\psi_i$, $\psi_i'$, it is possible to define a rotated basis $\psi_\pm \sim \psi_i \pm \psi_i'$ where $\psi_\pm$ are of definite parity.
So, if $[P, H] = 0$, then the Hilbert space can be spanned by real-valued $\psi_i(\vec x)$ that are eigenstates of both parity and energy.

Central potentials $V(x)$, e.g.~hydrogenic atoms, provide a familiar example. It is common to label the electronic states of fixed energy according to their angular quantum numbers $\ell, m$, i.e. $\psi_i = \ket{n \ell m}$, 
with $\psi_{n \ell m}(\vec x) \propto Y_\ell^m(\hat x)$ or $\psi_{n\ell m}(\vec k) \propto Y_\ell^m(\hat k)$.
In this case, real-valued parity eigenstates can be constructed via the linear combinations $Y_{\ell}^{m} \pm Y_\ell^{-m}$.

\paragraph{Effect of Parity on Form Factors:}

Having established a basis of eigenstates of $P$ and $H$, we move on to the scalar and vector form factors, $f_S(\vec q)$ and $\vec f_{1 \rightarrow 2}(\vec q)$. 
Because $\psi_{1,2}(\vec x)$ are real-valued with definite parity, the same is true for the product $\psi_2^\star(\vec x) \psi_1(\vec x)$.\footnote{Since the $\psi_i$ are real-valued, $\psi_i^\star = \psi_i$.} 
From \eqref{def:fourier}, $f_S(\vec q)$ is simply the Fourier transform of this function, or rather its complex conjugate: 
\begin{align}
f_S(- \vec q) &= \mathcal F[ \psi_2^\star \psi_1 ] = \mathcal F[ \psi_2 \psi_1 ] =  \int\! d^3x\, e^{-i \vec q \cdot \vec x} \, \psi_2(\vec x) \psi_1(\vec x),
\\
f_S(\vec q) &= f_S^\star(-\vec q)\, .
\label{eq:fSstarminus}
\end{align}
Noting that the Fourier transform commutes with parity,
\begin{align}
f_S(\vec q) &= P \cdot f_S(- \vec q) = \mathcal F[P \cdot \psi_2 \psi_1] \equiv \mathcal F[(\pm 1) \psi_2 \psi_1] \,,
\\
&\equiv \epsilon_{12} f_S( - \vec q)\, , 
\label{eq:parity_q_opp}
\end{align}
where $\epsilon_{12} = +1$ if the function $\psi_2 \psi_1$ is even, and $\epsilon_{12} = -1$ if it is odd. 
Applying \eqref{eq:fSstarminus} to this result, we see that 
\begin{align}
f_S^\star(\vec q) &= \epsilon_{12} f_S(\vec q)\, .
\label{eq:fSImRe}
\end{align}
Therefore, if $\psi_{1,2}$ have matching parity under central inversion ($\epsilon_{12}=+1$), then $f_S$ is purely real; if $\psi_{1,2}$ have opposite parities ($\epsilon_{12}=-1$), then $f_S$ is purely imaginary.

Next, consider the quantity $i \vec f_{1 \rightarrow 2}$,
\begin{align}
i m_e \vec f_{1 \rightarrow 2}(\vec q) &= \int\! d^3x\, e^{i \vec q \cdot \vec x} \psi_2^\star(\vec x) \cdot \nabla \psi_1(\vec x)
=\big(  \mathcal F[ \psi_2^\star \nabla \psi_1 ] \big)^\star
=\big(  \mathcal F[ \psi_2 \nabla \psi_1 ] \big)^\star .
\end{align}
The $\nabla$ operator flips the parity of the integrand, so 
\begin{align}
\label{eq:parityvec1}
P \cdot i m_e \vec f_{1 \rightarrow 2}(\vec q) &= 
\mathcal F[ P \cdot \psi_2 \nabla \psi_1]^\star= (- \epsilon_{12}) \mathcal F[ \psi_2 \nabla \psi_1]^\star 
= (- \epsilon_{12}) i m_e \vec f_{1 \rightarrow 2}(\vec q). 
\end{align}
For real-valued $\psi_{1,2}$, the parity operation $P \cdot \vec f_{1 \rightarrow 2}$ is equivalent to complex conjugation: 
\begin{align}
\label{eq:parityvec2}
-i m_e \vec f_{1 \rightarrow 2}^\star (\vec q) 
&= \mathcal F[ \psi_2 \nabla \psi_1 ] 
= i m_e \vec f_{1 \rightarrow 2}(- \vec q) ,
\end{align}
or more simply 
\begin{align}
\vec f_{1 \rightarrow 2}^\star(\vec q) &= - \vec f_{1 \rightarrow 2}(-\vec q)
= + \epsilon_{12}   \vec f_{1 \rightarrow 2}(\vec q). 
\label{eq:f12ImRe}
\end{align}
Again, we see here that $\epsilon_{12}=\pm 1$ for $\psi_2\psi_1$ even/odd, respectively.\footnote{In this section, we take the parity operator to act on the spatial coordinates. Eqns.~(\ref{eq:fSImRe}) and~(\ref{eq:f12ImRe}) also follow unchanged if the parity operator acts on the Hilbert space, but the commutation relations between parity and momentum introduce extra minus signs in the intermediate equations.} Therefore, $f_S$ and $\vec f_{1\rightarrow2}$ transform the same under parity, i.e. if $f_S$ is real, then $\vec f_{1 \rightarrow 2}$ is also real; if $f_S$ is imaginary, then so is $\vec f_{1 \rightarrow 2}$.

The fact that $f_S$ and $\vec f_{1 \rightarrow 2}$ in a parity-symmetric system are either purely real or purely imaginary has immediate, important consequences for the material response functions. 
First, parity symmetry enforces that $B_5$ vanishes.
From~\eqref{def:B5}, the only way to obtain a nonzero $B_5$ is if $\vec f_{1 \rightarrow 2}$ has both real and imaginary components,
\begin{align}
B_5 &= \frac{2 \vec q}{m_e} \cdot \left( \text{Re}(\vec f_{1 \rightarrow 2} ) \times \text{Im}(\vec f_{1 \rightarrow 2}) \right) ,
\label{eq:B5alt}
\end{align}
and \eqref{eq:f12ImRe} implies that either $\text{Re}(\vec f_{1\rightarrow 2}) = 0$ or $\text{Im}(\vec f_{1\rightarrow 2}) = 0$.

Second, parity symmetry implies that $\vec A$ is real-valued:
\begin{align}
\vec A^\star &= f_S^\star \vec f_{1 \rightarrow 2} = (\epsilon_{12} f_S) ( \epsilon_{12} \vec f_{1 \rightarrow 2}^\star) = \epsilon_{12}^2 \vec A = \vec A\, ,
\label{eq:parityAreal}
\end{align}
which in turn implies that $B_2, B_6,$ and $B_7$ are real-valued. This allows us to simplify several of the terms in~\eqref{eq:a01234B1234567}.
The $a_3$ component of $\overline{|\mathcal M_{1 \rightarrow 2}|^2}$  vanishes completely:
\begin{align}
 i a_3 \frac{q}{m_e} \cdot (\vec f_V \times \vec f_V^\star) &= a_3 \left[ B_5 - 2 \, \text{Im}\left( \vec v \cdot \vec B_7 \right) \right]
 = 0.
\label{eq:paritya3}
\end{align}
If $a_4$ is real-valued, then the $a_4$ component of $\overline{|\mathcal M_{1\rightarrow 2}|^2}$ also vanishes.
Finally, for real $B_2$, \eqref{eq:B124} becomes
\begin{align}
(B_2)^2 &=  B_4 B_1.
\end{align}
This modestly simplifies the $a_2$ term.
\paragraph{Matrix-element:}
In conclusion, the scattering rate in a parity-symmetric material can be described by just three linear combinations of the $B_{1 \ldots 3}$ and $\vec B_6$ material response functions, 
\begin{align}
\overline{|\mathcal M_{1 \rightarrow 2} |^2} &= a_0 | f_S|^2 + a_1|\vec f_V|^2 
+ \frac{a_2}{q^2/m_e^2} \left| \frac{\vec q}{m_e} \cdot \vec f_V \right|^2 
\\
&= a_0 B_1 
+ a_1 \left[ \left( v^2 - v_\text{min}^2  \right) B_1     - 2 \vec v \cdot \Aperp
 + B_3^\perp
 \right]
+ (a_1 + a_2)  B_H
\end{align}
This result applies to crystals composed of $P$-symmetric molecules, e.g.~\cite{Blanco:2019lrf,Blanco:2021hlm}, as well as semiconductors with $P$-symmetric lattices. If the Hamiltonian $H$ permits degenerate excited states, the cancellation between the $a_3$ and $a_4$ contributions only becomes apparent when summing over the degenerate final states: i.e.
\begin{align}
\sum_{f}\overline{|\mathcal M_{1 \rightarrow f} |^2} &=  \sum_{f} \left( a_0 | f_S|^2 + a_1|\vec f_V|^2 
+ \frac{a_2}{q^2/m_e^2} \left| \frac{\vec q}{m_e} \cdot \vec f_V \right|^2  \right) ,
\label{eq:a012}
\end{align}
summing over all $\psi_f$ satisfying $H \psi_f = E_2 \psi_f$.

\subsection{Isotropic Detectors} \label{sec:iso}
Another common symmetry found in materials is isotropy. 
A material may be isotropic because its $\psi_{1,2}$ wavefunctions are eigenstates of a spherically symmetric Hamiltonian, with $V(\vec x) = V(x)$. 
Alternatively, as in the cases of fluids and glasses, the material is isotropic only because its constituents are randomly aligned (e.g.~liquid molecular scintillators~\cite{Blanco:2019lrf}).  
The total scattering rate is then proportional to isotropic averages of the anisotropic material response functions. 
In this section, we explore how isotropy simplifies the scattering rate, even in chiral materials.\footnote{Chirality is rotation-invariant, so a liquid of chiral molecules retains its chirality.}

\paragraph{Rotation Operator:} Following  \eqref{eq:rateR}, 
an isotropic detector can be described as a collection of unit cells, each with a different orientation $\mathcal R$, sampled uniformly on the $SO(3) \equiv S^3/Z_2$ manifold. 
To begin, we review how the rotation operator acts on the wavefunctions $\psi_{1,2}$.
We take $\mathcal R$ to act on the detector, rather than the DM velocity distribution, so that $\mathcal R$ acts only on the wavefunctions $\psi_{1,2}$:
\begin{align}
\mathcal R \cdot \psi(\vec u) &= \psi(\mathcal R^{-1} \cdot \vec u)\,, 
\end{align}
for $\vec u = \vec x$ or $\vec u = \vec k$ in position or momentum space. 
However, by a change of variable, this is equivalent to acting on the $\vec q$ dependent form factors $f_S(\vec q)$ and $\vec f_{1 \rightarrow 2}(\vec q)$:
\begin{align}
\mathcal R \cdot f_S(\vec q) &= f_S(\mathcal R^{-1} \vec q ) , 
&
\mathcal R \cdot \vec f_{1 \rightarrow 2}(\vec q) &= [\mathcal R \vec f_{1 \rightarrow 2}](\mathcal R^{-1} \vec q )\, .
\end{align}
Rotations on a vector field $\vec f_{1 \rightarrow 2}$ change the direction of the vector, $\mathcal R \vec f$, as well as its dependence on its argument. 
After simplifying the dot products $\vec q \cdot \vec f(\vec q) \rightarrow \vec q \cdot \mathcal R \vec f(\mathcal R^{-1} \vec q) = (\mathcal R^{-1} \vec q) \cdot \vec f(\mathcal R^{-1} \vec q)$, 
all the scalar response functions $B_{1 \ldots 5}$ transform as:
\begin{align}
\mathcal R \cdot B_{i}(\vec q) &= B_i(\mathcal R^{-1} \vec q)\, .
\end{align}
\paragraph{Isotropic Averaging of Form Factors:} 
In a detector with $ N_\text{cell} \sim  N_A$ many randomly-aligned constituent particles or unit cells, it is an excellent approximation to ignore the $1/\sqrt{N_\text{cell}}$ statistical fluctuations away from spherical symmetry.
Each material response function can then be replaced by its isotropic average:
\begin{align}
\langle f(\vec q) \rangle \equiv \frac{1}{V_{SO(3)}} \int_{SO(3)}\!d\mathcal R \left( \mathcal R \cdot f(\vec q) \right)\, , 
\label{eq:BisoInt}
\end{align}
an equally weighted sum over all $\mathcal R \in SO(3)$, normalized by the volume $V_{SO(3)} = \frac{1}{2} 2\pi^2$.
If $f(\vec q) = f(q)$ is already spherically symmetric, then each $\mathcal R f = f$ acts trivially on the response functions.
We note that the isotropic average is a function only of $q = |\vec q|$:
\begin{align}
B^\text{iso}_{i}(q) &\equiv \langle B_i(\vec q) \rangle\,, 
\label{eq:Biso}
\end{align}
for $i = 1, \ldots, 5$. 

For the vectorial response functions $\vec B_{6,7}$, on the other hand, we must track how $\mathcal R$ acts on the vector field $\vec A(\vec q)$. 
Using the index notation with unit vectors $\hat e_i$, where repeated indices imply summation, $\vec A(\vec q) = A_i(q_j \hat e_j) \hat e_i$ transforms under rotations as 
\begin{align}
\mathcal R \cdot \vec A(\vec q) &=  A_{i}(q_j \mathcal R_{jb} \hat e_b) \, \mathcal R^{-1}_{ia} \hat e_a, 
\end{align}
where $\mathcal R^{-1}$ is the inverse of the $3\times 3$  matrix representation of the active rotation $\mathcal R \in SO(3)$. 
A generic $\vec A$ has a radial component parallel to $\vec q$, which we denote as $A^\parallel \, \hat q$, and a tangential component $\vec A^\perp$ satisfying $\vec q \cdot \vec A^\perp = 0$:
\begin{align}
\vec A(\vec q) &= A^\parallel(\vec q) \hat q + \vec A^\perp(\vec q)\,. 
\label{eq:Aperp}
\end{align}
In the isotropic average of $\vec A$, $\vec A_\text{iso}(q) = \langle \vec A(\vec q) \rangle$, the contributions from $\vec A^\perp$ vanish. 
This can be seen most simply by the requirement that $\langle \vec A^\perp\rangle$ should be rotationally invariant. 
If $\langle \vec A^\perp(\vec q) \rangle \neq 0$ at some point $\vec q$, then we can apply a $180^\circ$ degree rotation about the $\hat q$ axis:
\begin{align}
\mathcal R_\pi^{(\vec q)} \cdot \langle \vec A^\perp(\vec q) \rangle = - \langle \vec A^\perp(\vec q) \rangle = 0\,.
\end{align}
This leaves $\vec q$ fixed, but rotates $\vec A^\perp \rightarrow - \vec A^\perp$. So, a rotationally invariant $\vec A^\perp$ must  be zero.\footnote{This is a corollary of the fuzzy ball theorem for vector fields defined on the $S^2$ sphere: if $\vec A^\perp$ is continuous, then there must be at least two points at which $\vec A^\perp(\vec q) = 0$. Requiring that $\vec A^\perp_\text{iso}(\vec q)$ is also rotationally invariant implies that $\vec A^\perp_\text{iso} = 0$ everywhere.
}
In the isotropic average, then, 
\begin{align}
\vec B_6^\text{iso}(q) &\equiv \vec A_\text{iso}(q) = A^{\parallel}_{\rm{iso}}(q)  \, \hat q
\end{align}
can include only a radial component, and 
\begin{align}
\vec B_7^\text{iso}(q) &= \frac{1}{m_e} \langle \vec q \times \vec A(\vec q) \rangle = \left\langle \frac{\vec q}{m_e} \times \left( A^\parallel(\vec q) \hat q\right) \right\rangle = 0
\end{align}
vanishes identically. 
While $\vec B_6^\text{iso}$ does not generally vanish, it is a simple function of  $\vec q \, B_2(\vec q)$: 
\begin{align}
\vec B_6^\text{iso}=\vec A_\text{iso}( q) &= \frac{\vec q}{q^2} \left( \vec q \cdot \vec A_\text{iso}(q) \right) = \frac{\vec q}{m_e} \frac{m_e^2}{q^2} B_2^\text{iso}(q)\, ,
\end{align}
so that 
\begin{align}
\vec v \cdot \vec B_6^\text{iso}(\vec q) &= \frac{\vec v \cdot \vec q}{m_e} \frac{m_e^2}{q^2}  B_2^\text{iso}(q)\, . 
\end{align}
So, the combination of $B_2$ and $\vec B_6$ that appears in \eqref{eq:a01234B1234567} can be simplified to:
\begin{align}
\left[
\frac{m_e}{\mu_{\chi e} } B_2 - 2 \vec v \cdot \text{Re}\, \vec B_6 
\right]^\text{iso} &= - \frac{2 m_e}{q^2} (\vec q \cdot \vzero) \text{Re}\, B_2^\text{iso}(q) +  \frac{m_e}{\mu_{\chi e}}\text{Im} (B_2^{\rm iso})\,. 
\end{align}
Although $\langle \vec A^\perp \rangle = 0$ in the isotropic average, this does not mean that $\vec A^\perp$ can be completely ignored: $B_3$, for example, is quadratic in $\vec f_{1 \rightarrow 2} \propto \vec A$, and 
\begin{align}
|\vec A |^2 &= \left| \hat q \cdot \vec A  \right|^2 + (\vec A^\perp)^\star \cdot \vec A^\perp, 
\\
\langle B_1 B_3 \rangle &= \left\langle |A^{||}|^2 + |\vec A^\perp|^2 \right\rangle\, .
\end{align}
Finally, let us consider $\langle B_5(\vec q) \rangle$.
Note that $B_5$ can be decomposed as
\begin{align}
B_5^\text{iso}(q) &=   \left\langle \frac{2 \vec q}{m_e} \cdot \text{Re}(\vec f_{1 \rightarrow 2} ) \times \text{Im}(\vec f_{1 \rightarrow 2}) \right\rangle .
\end{align}
In isotropic configurations of anisotropic unit cells, there can be tangential components of $\vec f_{1 \rightarrow 2}(\vec q)$, and in a chiral system there is no requirement for $\text{Re}(\vec f_{1 \rightarrow 2}) \times \text{Im}(\vec f_{1 \rightarrow 2})$ to vanish. 
Therefore, $B_5$ can contribute to isotropic-but-chiral detector materials. 

\paragraph{Matrix Element:} After the isotropic average, then, the spin averaged squared amplitude is:
\begin{align}
\left\langle \overline{|\mathcal M_{1 \rightarrow 2} |^2} \right\rangle &= a_0 B_1^\text{iso}(q) 
+ a_1 \left[ |\vzero|^2 B_1^\text{iso}(q) - \frac{2 m_e^2}{q^2} \frac{\vec q \cdot \vzero }{m_e} \text{Re}\,B_2^\text{iso}(q) + B_3^\text{iso}(q) \right]
\nonumber\\&~~
+ a_2 \frac{m_e^2}{q^2} \left[ \left( \frac{\vec q \cdot \vzero}{m_e} \right)^2 B_1^\text{iso}(q)  - 2 \left( \frac{\vec q \cdot \vzero}{m_e} \right) \text{Re}\,  B_2^\text{iso}(q)  + B_4^\text{iso}(q)  \right] 
\nonumber\\ &~~
+ a_3 B_5^\text{iso}(q) 
- 2 \, \text{Re}\left( a_4 \right) \text{Im}\,B_2^\text{iso}(q)\,  .
\label{eq:isochiral}
\end{align}
Recall that $a_i(q)$ are functions of $q = | \vec q|$, but not $\vec v$. 
In a parity-symmetric, rotationally invariant detector, 
\begin{align}
\left\langle \overline{|\mathcal M_{1 \rightarrow 2} |^2} \right\rangle &= a_0 B_1^\text{iso}(q) 
+ a_1 \left[ |\vzero|^2 B_1^\text{iso}(q) - \frac{2 m_e^2}{q^2} \frac{\vec q \cdot \vzero }{m_e} B_2^\text{iso}(q) + B_3^\text{iso}(q) \right]
\nonumber\\&~~
+ a_2 \frac{m_e^2}{q^2} \left[ \left( \frac{\vec q \cdot \vzero}{m_e} \right)^2 B_1^\text{iso}(q)  - 2 \left( \frac{\vec q \cdot \vzero}{m_e} \right)   B_2^\text{iso}(q)  + B_4^\text{iso}(q)  \right]\, .
\end{align}
Although $B_4 = B_2^2 / B_1$, it is not necessarily true that $B_4^\text{iso} = (B_2^\text{iso})^2 / B_1^\text{iso}$. That would only be guaranteed if $B_2(\vec q)$ and $B_1(\vec q)$ are both spherically symmetric, which is a much more restrictive assumption.

\paragraph{Discrete Rotational Symmetries:}
Many crystal groups include one or more discrete rotational symmetries, e.g.~under rotations of $180^\circ$ about some axis $\hat b$. 
While the linear terms in $\vec A^\perp$ (i.e.~$\vec B_{6,7}$) cannot \emph{generally} be ignored, we can see immediately that $\vec A^\perp(q \hat b) = 0$ for $\vec q = q \hat b$ along the axis of symmetry. 
In the case of degenerate $E_2$ final states, this argument assumes that the energy eigenstates are labeled using eigenstates of the discrete rotation operator; otherwise, the cancellation of $A^\perp(q \hat b)$ arises later in the calculation, in the sum over degenerate states. 
$\vec A^\perp$ still needs to be evaluated for generic off-axis values of $\vec q$, however. 

\begin{table}[]
    \centering
    \begin{tabular}{c||c|c|c|c|c|c|c}
         {\bf material property }& $B_1$ & $B_2$&$B_3$&$B_4$&$B_5$&$\vec B_6$&$\vec B_7$ \\
         \hline
         parity symmetry $[H,P]=0$ &\ding{52} & Real & \ding{52} & $B_1, B_2$ & \ding{56} & Real & \ding{56}$^\star$ \\
         isotropy $V(\vec q)=V(q)$& \ding{52} & \ding{52} &\ding{52}&\ding{52}&\ding{52}& $B_2$&\ding{56}\\
         $e^-$ adiabaticity $\hat H_1=\hat H_2$ & \ding{52} & $B_1$ & \ding{52} & $B_1$ & \ding{52} & Real & Imaginary\\
         \hline
    \end{tabular}
    \caption{Summary of which material response functions contribute to the rate under the symmetries or simplifying regimes discussed in Sec.~\ref{sec:simplified}. ($\star$) Under parity, $\vec B_7$ must be real, but ${\rm Re}(\vec B_7)$ does not appear in the rate (Eq.~\ref{eq:a01234B1234567}) and therefore $\vec B_7$ never contributes.}
    \label{tab:materialresponse}
\end{table}

\subsection{Electronic Adiabaticity} \label{sec:adiabatic}
In this section, we will derive the relationships between $B_1$ and some of the $B_{i \geq 2}$ response functions for the class of physical systems where $\psi_1$ and $\psi_2$ are eigenstates of a Hamiltonian that acts on position space $\psi_{1,2}(\vec x)$ as 
\begin{align}
H_0 = - \frac{\nabla^2}{2 m} + V(\vec x)\, ,
\label{eq:Heasy}
\end{align}
with a potential $V(\vec x)$ that depends only $\vec x$, and not its conjugate momentum. 
Here, depending on the system, $m$ can represent i.e. the reduced or effective mass of the system. We will show that when $V(\vec x)$ is the same for the initial and final states, we can use our derivation to replace all the appearances of $\vec q \cdot \vec f_{1 \rightarrow 2}$ with simply $f_S$. 
For $\vec q$ and $\vec v$ satisfying the energy conservation condition 
\begin{equation}
\label{eq:energycons}
    \vec q \cdot \vec v = E_2 - E_1 + q^2/(2m_\chi)\, ,
\end{equation} 
we show that the terms multiplying $a_2$ and $a_4$ in \eqref{eq:a01234B1234567} vanish.
We refer to this special case with unchanging $V_1(\vec x) = V_2(\vec x)$ as ``electronic adiabaticity,'' because it implies that the Coulomb potential from the atomic nuclei \emph{and} the other electrons in the system is unchanged by the DM-induced excitation of one electron.
Aside from single-electron systems like the hydrogen atom or the $H_2^+$ ion, examples of approximate $e^-$ adiabaticity include the many atomic or molecular systems where the excitation above the ground state is well described by a single pair of Hartree--Fock eigenstates~\cite{Hartree_1928,RevModPhys.23.69}, as in the first excited state in many aromatic organic molecules~\cite{Blanco:2019lrf, Blanco:2021hlm}.

Section~\ref{sec:IBP} begins without assuming that $V(\vec x)$ has any spatial symmetries, 
showing that the spin-averaged amplitude squared in \eqref{eq:a01234B1234567} reduces to just the $a_0$, $a_1$, and $a_3$ terms. 
In Section~\ref{sec:IBPsymmetry}, we specialize to parity-symmetric and isotropic systems, using our results to further simplify the dependence on the material response functions.

\subsubsection{Integration By Parts} \label{sec:IBP}
To simplify the presentation of our derivation, we define a symmetric vector form factor $\vec f_\leftrightarrow$:
\begin{align}
\vec f_{\leftrightarrow}(\vec q) &\equiv \frac{-i}{m} \int \! d^3 x\, e^{i \vec q \cdot \vec x} \left( \psi_2^\star \nablaLR \psi_1 \right)\, ,
\label{def:symmf}
\end{align}
where 
\begin{align}
\psi_2^\star \nablaLR \psi_1 \equiv \psi_2^\star \nabla \psi_1 - \psi_1 \nabla \psi_2^\star\, .
\end{align}
Up to a surface integral, this $\vec f_{\leftrightarrow}$ is related to the original $\vec f_{1\to 2}$ and $f_S$ via integration by parts:
\begin{align}
\vec f_{\leftrightarrow}(\vec q) &= \vec f_{1\to 2}(\vec q)+ \frac{i}{m} \int\! d^3x\, e^{i \vec q \cdot \vec x} \psi_1 \nabla \psi_2^\star
\\
&= 2\, \vec f_{1\to 2}(\vec q) + \frac{\vec q}{m} f_S(\vec q)  + \frac{i}{m} \int_{\partial V}\! d\Gamma\, \hatn e^{i \vec q \cdot \vec x} \psi_2^\star \psi_1\, ,
\label{eq:fLRibp}
\end{align}
where $\hatn$ is the unit normal vector to the surface $\partial V$ on the boundary of the integration volume, and $d\Gamma$ is an infinitesimal surface element on $\partial V$. 

In isolated systems (e.g.~nuclei, atoms, molecules) the $\int \! d^3 x$ integral usually extends to $|\vec x| \rightarrow \infty$, 
and so the boundary integral vanishes because $\psi_1(\vec x \rightarrow \infty) = 0$ for the ground state. 
For crystals, $\int\!d^3 x$ may be applied instead to a single unit cell, with $\psi_{1,2}(\vec x)$ given periodic boundary conditions at the finite surface $\partial V$. 
In this example, assuming $\psi_{1,2}$ are continuous and topologically trivial, the surface integrals cancel pairwise: neighboring unit cells have matching $e^{i \vec q \cdot \vec x} \psi_2^\star(\vec x) \psi_1(\vec x)$ at the boundary, with opposite-sign $\pm \hat n$ normal vectors. 
For completeness, we include the boundary terms in the following results. However, as long as there are no topologically nontrivial edge states, we expect these surface integrals to vanish.

\paragraph{Simplifying $B_2$ and $B_4$:} 
We note that $\vec q \cdot \vec f_\leftrightarrow$ can be written as an integral involving a gradient of $e^{i \vec q \cdot \vec x}$: 
\begin{align}
- m \vec q \cdot \vec f_\leftrightarrow
&= \int\! d^3 x \, e^{i \vec q \cdot \vec x} i \vec q \cdot \left( \psi_2^\star \nablaLR \psi_1 \right)
= \int \! d^3 x\, \nabla(e^{i \vec q \cdot \vec x}) \cdot  \left( \psi_2^\star \nablaLR \psi_1 \right)\, .
\end{align}
Integrating this expression by parts, we find:
\begin{align}
 \vec q \cdot \vec f_{\leftrightarrow}(\vec q) &= - \int_{\partial V}\! d\Gamma \, e^{i \vec q \cdot \vec x} \frac{\hatn}{m} \cdot ( \psi_2^\star \nablaLR \psi_1)
+ 2\int\! d^3 x \, e^{i \vec q \cdot \vec x} \frac{\psi_2^\star \nabla^2  \psi_1 - \psi_1 \nabla^2 \psi_2^\star  }{2 m}\, .
\label{eq:qdotLRf}
\end{align}
If $\psi_{1,2}$ are both solutions to second order differential equations of the form \eqref{eq:Heasy}, i.e.~$H_i \psi_i = E_i \psi_i$ with $ \nabla^2/(2 m) = -H_i + V_i(\vec x)$ for potentials $V_{i = 1,2}$, 
then 
\begin{align}
 \vec q \cdot \vec f_{\leftrightarrow}(\vec q) &= - \int_{\partial V}\! d\Gamma \, e^{i \vec q \cdot \vec x} \frac{\hatn}{m} \cdot ( \psi_2^\star \nablaLR \psi_1)
+ 2\int\! d^3 x \, e^{i \vec q \cdot \vec x} \left[\psi_2^\star (V_1-E_1) \psi_1 + \psi_1 (E_2 - V_2) \psi_2^\star \right]
\end{align}
The potentials $V_{1,2}(\vec x)$ commute with the position-space wavefunctions, leaving only 
\begin{align}
\vec q \cdot \vec f_{\leftrightarrow}(\vec q) &=
2 (E_2 - E_1) \, f_S(\vec q) - \int_{\partial V}\! d\Gamma \, e^{i \vec q \cdot \vec x} \frac{\hatn}{m} \cdot ( \psi_2^\star \nablaLR \psi_1)
- 2 \langle \psi_2 | e^{i \vec q \cdot \vec x } \, \delta V | \psi_1 \rangle\, , 
\end{align}
where for the generic case $V_2 \neq V_1$ we define:
\begin{align}
\langle \psi_2 | e^{i \vec q \cdot \vec x } \, \delta V | \psi_1 \rangle
&\equiv \int\! d^3x \, e^{i \vec q \cdot \vec x} \big(V_2(\vec x) - V_1(\vec x) \big) \psi_2^\star \psi_1\,.
\label{eq:intdV}
\end{align}
For electronically adiabatic systems, $\langle \psi_2 | e^{i \vec q \cdot \vec x } \, \delta V | \psi_1 \rangle = 0$. 

Like the boundary term in \eqref{eq:fLRibp}, the $(\psi_2^\star \nablaLR \psi_1)$ part of the integrand vanishes for isolated systems, as $\psi_1$ and $\nabla \psi_1$ both vanish in the $|\vec x| \rightarrow \infty$ limit. 
For periodic crystals, the $\pm \vec n$ cancellation from adjacent unit cells also applies to the new integrand, as long as $(\psi_2^\star \nablaLR \psi_1)$ is continuous at the boundary of the unit cell (i.e.~assuming there is no infinitely dense surface charge distribution at $\partial V$). Topologically nontrivial wavefunctions at the physical surface of the detector volume could in principle generate a nonzero boundary term.

With \eqref{eq:fLRibp}, it is easy to convert $\vec q \cdot \vec f_\leftrightarrow = 2 E \, f_S$ into a similar expression for $\vec q \cdot \vec f_{1 \rightarrow 2}$, 
\begin{align}
\vec q \cdot \vec f_{1 \rightarrow 2} &= \left( E_2 - E_1 - \frac{q^2}{2 m} \right) f_S
- \frac{1}{2 m} \int_{\partial V} \! d\Gamma \, e^{i \vec q \cdot \vec x} \psi_2^\star \left( \vec n \cdot \nablaLR + i \vec q \cdot \vec n \right) \psi_1\, ,
\end{align}
which can in turn be used to write $B_2$ and $B_4$ in terms of the simpler $B_1$ response function.
Using the usual arguments to drop the boundary terms, and \eqref{eq:energycons} we find:
\begin{align}
\vec q \cdot \vec f_{1 \rightarrow 2} &= (\vec q \cdot \vzero) f_S, 
&
\vec q \cdot \vec A & = (\vec q \cdot \vzero) B_1,
\label{eq:qdotA}
\\
B_2 &= \frac{\vec q \cdot \vzero}{m}  B_1, 
&
B_4 &= \left( \frac{\vec q \cdot \vzero}{m} \right)^2 B_1\,.
\end{align}
Notice that $B_2$ is now explicitly real-valued.  

For generic anisotropic detectors, the part of $\vec f_{1\to 2}$ that is perpendicular to $\vec q$ (i.e.~$\vec q \times \vec f$) is not simplified by this trick, so we still expect to need $B_3$, $B_5$, $\vec B_6$, and $\vec B_7$. 
This is where the organization scheme of Ref.~\cite{Liang:2024ecw} becomes particularly convenient. Rather than $\vec q \cdot \vec f_{1\to 2}$, consider instead the velocity-dependent vector form factor, $\vec q \cdot \vec f_V$: 
\begin{align}
\vec q \cdot \vec f_V(\vec q, \vec v) &= -\left( E_2 - E_1 + \frac{ q^2}{2 m_\chi} - \vec q \cdot \vec v  \right) f_S(\vec q)\,.
\label{eq:noqdotfV}
\end{align}
This combination of $E$, $q^2$, and $\vec q \cdot \vec v$ is the same one that appears in the argument of the energy-conserving $\delta$ function in \eqref{eq:rate}. Consequently, for on-shell momenta $\vec q$, for any non-relativistic $\vec v$, we find 
\begin{align}
\vec q \cdot \vec f_V \rightarrow 0\,,
\end{align}
up to some surface integrals that vanish under the typical assumptions.

As a verification of our results, we test \eqref{eq:qdotA}
in two single particle quantum systems where we know \textit{a priori} that $\delta V = 0$. These examples are provided in Appendix~\ref{app:explicit}. First, as a crude approximation of a crystal, we take $\psi_{1,2}$ to be energy eigenstates of a particle in an infinite rectangular well, with arbitrary box dimensions $(L_x, L_y, L_z)$. Next, we consider an isolated hydrogen atom, with $\Psi_2$ any of the excited states. 
Both of these examples permit analytic solutions for $f_S$ and $\vec f_{1 \rightarrow 2}$, as well as $E_1$ and $E_2$.
As expected, we find:
\begin{align}
\vec q \cdot \vec f_{1 \rightarrow 2}(\vec q) = \left( E_2 - E_1 - \frac{q^2}{2m} \right) f_S(\vec q) \, .
\end{align}

\paragraph{Simplifying $B_3$ and $\vec B_6$:} 
For the remainder of our discussion, is helpful to separate the $\vec f_{1 \rightarrow 2}(\vec q)$ form factor into a radial part, $\propto \hat q$, and a tangential part $\vec f_{1 \rightarrow 2}^\perp$, as in \eqref{eq:altB1234567}: 
\begin{align}
\vec f_{1 \rightarrow 2} &= (\hat q \cdot \vec f_{1 \rightarrow 2}) \hat q + \vec f_{1 \rightarrow 2}^\perp \,.
\end{align}
The radial part, though not $\vec f_{1 \rightarrow 2}^\perp$, is related to $f_S$ via integration by parts: 
\begin{align}
\vec f_{1 \rightarrow 2}(\vec q) &=  \frac{\vec q \cdot \vzero }{q} f_S(\vec q) \hat q + \vec f_{1 \rightarrow 2}^\perp (\vec q),
\\
\vec A (\vec q) = f_S(\vec q) \vec f^\star_{1 \rightarrow 2}(\vec q) &= \frac{\vec q \cdot \vzero}{q} B_1(\vec q) \, \hat q + \vec A^\perp(\vec q)\,, 
\end{align}
where $\vec A^\perp = f_S (\vec f_{1 \rightarrow 2}^\perp )^\star$. 
So, $B_3$ and $\vec B_6$ separate into a term proportional to $B_1$ plus a $B_{3,6}^\perp$ term generated only by $\vec f_{1 \rightarrow 2}^\perp$ or $\vec A^\perp$:
\begin{align}
\label{eq:B3_simplify}
B_3 &= \left( \frac{\vec q \cdot \vzero }{q} \right)^2 B_1 + B_3^\perp, 
&
B_3^\perp &\equiv \left| \vec f_{1 \rightarrow 2}^\perp \right|^2,
\\
\label{eq:B6_simplify}
\vec v \cdot \vec B_6 &= \frac{\vec v \cdot \vec q}{q} \frac{\vec q \cdot \vzero}{q} B_1 + \vec v \cdot \vec B_6^\perp, 
&
\vec B_6^\perp &\equiv 
 \vec A^\perp\,.
\end{align}
Note that the magnitudes of $B_6^\perp$ and $B_3^\perp$ are related: 
\begin{align}
\left| \vec B_6^\perp \right|^2 &= B_1 B_3^\perp\,.
\end{align}
Because $B_5$ and $\vec B_7$ are independent of the radial part of $\vec f_{1\rightarrow 2}$,
\begin{align}
B_5 &= \frac{2 \vec q}{m_e} \cdot \left( \text{Re}(\vec f_{1 \rightarrow 2}^\perp) \times \text{Im}(\vec f_{1 \rightarrow 2}^\perp) \right) , 
&
\vec B_7 &= \frac{\vec q }{m_e} \times \vec A^\perp\, ,
\end{align}
they are not simplified by the integration by parts trick.

\paragraph{Matrix Element:}
With the identification that $|\vzero|^2 - (\hat q \cdot \vzero)^2 = v^2 - v_\text{min}^2(q)$, we can write the simplified spin-averaged squared amplitude for a system where $\psi_{1,2}$ are eigenstates of the same Hamiltonian:
\begin{align}
\overline{|\mathcal M_{1 \rightarrow 2} |^2} &= a_0 B_1 
+ a_1 \left[ \left( v^2 - v_\text{min}^2 \right) B_1 + B_3^\perp - 2\, \text{Re}\left( \vec v \cdot \vec B_6^\perp \right) \right] 
+ a_3 \left[ B_5 - 2 \, \text{Im}\left( \vec v \cdot \vec B_7 \right) \right]\, .
\label{eq:a013}
\end{align}
From ~\eqref{eq:B3_simplify} and \eqref{eq:B6_simplify}, we see that $B_3^{||} = (\hat{q}\cdot\vec{v}_0^{\perp})^2 B_1$ and $B_6^{||} = (\hat{q}\cdot \vec{v}_0^\perp)B_1$, 
\begin{align}
B_H = (\hat{q}\cdot \vec{v}_0^\perp)^2 B_1 - 2 (\hat{q}\cdot \vec{v}_0^\perp)B_6^{||} + B_3^{||} \longrightarrow 0 \, .
\end{align}
The $a_1 B_1$ term is now explicitly nonnegative, vanishing when $v \rightarrow v_\text{min}(q)$ approaches the kinematic threshold for exciting the $E_2$ state.

\paragraph{Multiparticle Wavefunctions and Non-Adiabacity:}
In a system with multiple electrons, one generally expects a change to one electronic wavefunction to have some impact on the other electrons in the system. When the wavefunctions of the other electrons change, the Coulomb potential $V_2(\vec x)$ acting on $\psi_2$ changes, and $V_2 \neq V_1$ for the $e^-$ scattered by the DM. 
Even so, there are many cases where DM--$e^-$ scattering is well approximated by $V_1 \approx V_2$.
Precise \textit{ab initio} calculations, e.g.~post-Hartree--Fock methods or time-dependent density functional theory, can reveal whether or not $V_1 = V_2$ is a good approximation for a given system.

If $\psi_2$ and $\psi_1$ are eigenstates of \emph{different} Hamiltonians, but with the same effective particle mass $m$, then:
\begin{align}
- \frac{\nabla^2}{2m } &= H_1 - V_1 = H_2 - V_2\,,
\end{align}
with $V_1 \neq V_2$. We no longer expect perfect cancellation in $\vec q \cdot \vec f_V$. Instead, continuing to drop the $\int_{\partial V}$ surface integrals
and assuming the kinematics of elastic DM,
integration by parts gives us:
\begin{align}
\vec q \cdot \vec f_{1 \rightarrow 2} &= (\vec q \cdot \vzero) f_S  
-   \langle \psi_2 | e^{i \vec q \cdot \vec x} \,  \delta V | \psi_1 \rangle ,
\end{align}
for the $\langle \psi_2 | e^{i \vec q \cdot \vec x} \,  \delta V | \psi_1 \rangle$ defined in \eqref{eq:intdV}. 
For the velocity-dependent vector form factor, we find:
\begin{align}
\vec q \cdot \vec f_V &= \langle \psi_2 | e^{i \vec q \cdot \vec x } \, \delta V | \psi_1 \rangle\, .
\end{align}

A nonzero effective $\delta V$ generally arises in electronic systems whenever multiparticle effects are important. 
For example, in a system like trans-stilbene, the lowest energy transition is often dominated by one single particle process~\cite{ting1971,Blanco:2021hlm}. However, the higher-energy transitions can receive substantial contributions from multiple pairs of Hartree-Fock eigenstates and therefore it is not guaranteed that $\delta V \approx 0$ is a good approximation for these transitions.
Other examples where $\delta V$ cannot be neglected include non-adiabatic Migdal scattering in molecules~\cite{Blanco:2022pkt} and crystals~\cite{Esposito:2025iry}, as well as systems where exciton-hole interactions are important such as nanorods and quantum dots. 
In these systems, the initial and final states have somewhat different arrangements of the atomic nuclei, and the two electronic states have appreciably different Coulomb potentials.

\subsubsection{Inelastic DM} 
In the case of inelastic DM with a mass splitting $\delta$, the kinematic constraint relating $E_2-E_1$ and $\vec{q}$ is modified to
\begin{equation}
\label{eq:deltaE_inel}
    E_2-E_1 = - m_\chi \epsilon + \frac{\epsilon}{1+\epsilon} \frac{1}{2}m_\chi v^2 + \frac{\vec{q}\cdot\vec{v}}{1+\epsilon} - \frac{q^2}{2m_\chi (1+\epsilon)} \, ,
\end{equation}
where $\epsilon = \delta/m_\chi$. Applying these modified kinematics to \eqref{eq:noqdotfV}, we find instead:
\begin{equation}
    \vec{q}\cdot\vec{f}_V =  \left (  \frac{\epsilon}{1+\epsilon} \vec{q}\cdot \vec{v} - \frac{q^2}{2m_\chi}\frac{\epsilon}{1+\epsilon} + m_\chi \epsilon - \frac{\epsilon}{1+\epsilon} \frac{1}{2}m_\chi v^2\right ) f_S + \langle \psi_2 | e^{i \vec q \cdot \vec x } \, \delta V | \psi_1 \rangle .
\end{equation}
If $\psi_1$ is the ground state of the SM system, the necessary energy for the $m_\chi \rightarrow m_\chi + \delta$ transition must come from the DM kinetic energy, $\propto v^2 m_\chi$, with $v^2 \sim 10^{-6}$ for the galactic DM halo. 
Expanding to linear order in $v^2$ and $\epsilon \lesssim 10^{-5}$, then, 
\begin{equation}
    \vec{q}\cdot\vec{f}_V \simeq \epsilon \left [  m_\chi - \frac{q^2}{2m_\chi} + \vec{q}\cdot\vec{v}    \right ] f_S + \langle \psi_2 | e^{i \vec q \cdot \vec x } \, \delta V | \psi_1 \rangle + \mathcal O(\epsilon^2, \epsilon v^2).
\end{equation}

\subsubsection{Adiabaticity, Parity, and Isotropy} \label{sec:IBPsymmetry}

Section~\ref{sec:IBP} has shown that when $H_0 \psi_{1,2} = E_{1,2} \psi_{1,2}$ share the same Hamiltonian, the squared matrix element (\eqref{eq:a013}) is substantially simplified, compared to the generic spin-dependent case.
When this approximation can be applied to spherical or parity-symmetric  systems, the complexity is reduced even further.

If parity (under central inversion, $\vec x \rightarrow - \vec x$) is a good symmetry of the detector material, 
however, then Section~\ref{sec:parity} has shown that $\vec A = \vec B_6$ is real-valued.
In this case, the two  contributions to $a_3$ vanish separately, leaving only $a_0$ and $a_1$:
\begin{align}
\overline{|\mathcal M_{1 \rightarrow 2} |^2} &= a_0 B_1 
+ a_1 \left[ \left( v^2 - v_\text{min}^2 \right) B_1 + B_3^\perp - 2 \left( \vec v \cdot \vec B_6^\perp \right) \right]\, ,
\end{align}
for a parity-symmetric, electronically adiabatic detector.
If $H_0$ admits degenerate excited states, $H_0 \psi_2 = E_2 \psi_2$ and $H_0 \psi_2' = E_2 \psi_2'$, then this result holds as long as we label the final states using a basis where $\psi_2$ and $\psi_2'$ have definite parity. 
Otherwise, the cancellation of the $a_3$ term is only apparent in the sum over degenerate final states, $\sum_f \overline{|\mathcal M_{1 \rightarrow f} |^2} $.

Alternatively, consider the special case of an isotropic detector. As usual, we do not require that $V(\vec x) = V(x)$ itself is spherically symmetric, 
only that the material itself has no preferred directions, as in a fluid or glass. 
In such a system the scattering rate depends only on the isotropic average of $\overline{|\mathcal M_{1 \rightarrow 2}|^2}$, 
as in \eqref{eq:isochiral}.
Applying the rotational average to \eqref{eq:a013}, we find:
\begin{align}
\left\langle\overline{|\mathcal M_{1 \rightarrow 2} |^2} \right\rangle&= \Big( a_0 + a_1\left( v^2 - v_\text{min}^2 \right) \Big) B_1^\text{iso}
+ a_1 \left\langle B_3^{\perp} \right\rangle 
+ a_3  \left\langle B_5 \right\rangle\,, 
\end{align}
for an isotropic, possibly chiral, electronically adiabatic detector. Finally, if we assume that our adiabatic, isotropic system is also nonchiral:
\begin{align}
\left\langle\overline{|\mathcal M_{1 \rightarrow 2} |^2} \right\rangle&= \Big( a_0 + a_1\left( v^2 - v_\text{min}^2 \right) \Big) B_1^\text{iso}
+ a_1 \left\langle B_3^\perp \right\rangle\, .
\end{align}
Our assumption that $\psi_{1,2}$ are eigenstates of the same Hamiltonian has substantially simplified the spin-dependent form factor structure for isotropic systems, beyond any results in the literature. 

If the detector is isotropic not because it is a fluid or glass, but because its constituent particles or unit cells are themselves spherically symmetric, then $B_3^\perp \rightarrow 0$. 
In this simplest of $e^-$-adiabatic cases, the generic spin-dependent scattering rate depends only on the ``spin-independent'' material form factor $B_1(|\vec q|)$, 
\begin{align}
\left\langle\overline{|\mathcal M_{1 \rightarrow 2} |^2} \right\rangle& \longrightarrow \Big( a_0 + a_1\left( v^2 - v_\text{min}^2 \right) \Big) B_1^\text{iso}\, ,
\end{align}
up to corrections proportional to $\epsilon = \delta/m_\chi$ in the case of inelastic DM. 
Notably, this simplification applies to all of the spin-dependent operators shown in Figure~\ref{fig:tikz}, including the couplings to the electron spin, regardless of whether the DM is spin-0, spin-$\frac{1}{2}$ or spin-1.

\section{Case Studies} \label{sec:casestudies}
In this section, we give explicit examples of the coefficients $a_{0,\ldots,4}$ as in \eqref{eq:amp2Liang} as functions of the Wilson coefficients $c_i$ generated by three different UV models: anapole DM, Majorana DM, and scalar (inelastic) DM. The Wilson coefficients $c_i$ are those of the effective non-relativistic operators $\mathcal{O}_1,\ldots,\mathcal{O}_{15}$ defined in Table \ref{tab:NRops}. Our results are summarized in Table \ref{tab:a_coeffs}.
In Figure \ref{fig:tikz}, we show the effective operators that are generated from scalar and fermionic DM, as well as which $a_{0,\ldots,4}$ coefficients contribute to the response functions in \eqref{eq:aEtcB0}.

\paragraph{Kinematics:} Throughout the following case studies, we define $p^\mu$ and $p^{\prime \mu}$ as the incoming and outgoing four-momentum of the DM and $k^\mu$ and $k^{\prime\mu}$ as the incoming and outgoing four-momentum of the electron. In the center of momentum frame, we calculate the following kinematic quantities at linear order in $\vperp$:
\begin{align}
    \vec{p} &=-\vec{k} = \mu_{\chi e} \vperp + \frac{\vec{q}}{2}\\
    \vec{p}\,^\prime &= \vec{k}' =\mu_{\chi e} \vperp- \frac{\vec{q}}{2}\\
    p^0 &= E_p \simeq m_\chi \sqrt{\beta} + \frac{1}{\sqrt{\beta}}\frac{\mu_{\chi e}}{2m_\chi} (\vec{q}\cdot \vperp)\\
    p^{\prime 0} &= E_{p^\prime} \simeq m_\chi \sqrt{\beta}  - \frac{1}{\sqrt{\beta}}\frac{\mu_{\chi e}}{2m_\chi} (\vec{q}\cdot \vperp)
\end{align}
where 
\begin{equation}
\label{eq:beta_kin}
    \beta \equiv 1+\frac{q^2}{4m_\chi^2}\, .
\end{equation}

\subsection{Anapole Dark Matter}
Consider as an example the Lagrangian for spin-1/2 DM coupled to a massive vector through the anapole interaction
\begin{align}
    \mathcal{L}\supset
    & \frac{g_\chi}{4m_\chi^2}\bar{\chi}\gamma^\mu\gamma^5\chi\partial^\nu F'_{\mu\nu}+g_e\bar{e}\gamma^\mu eA'_\mu\,.
\end{align}
In addition to  the dark sector fermion $\chi$ (with mass $m_\chi$), we include a vector-like mediator $A'_\mu$, with mass $m_{A'}$ and field strength tensor $F'_{\mu\nu}=\partial_\mu A'_\nu-\partial_\nu A'_\mu$. 
This DM candidate interacts through its anapole moment under a spontaneously broken dark $U(1)$ gauge group. 
For a DM candidate with an anapole moment under the SM electromagnetism, simply set $m_{A'} \rightarrow 0$ and replace $g_e$ with the electric charge of an electron, $g_e \rightarrow - e$. 

We start by finding the amplitude in the Feynman gauge for the relevant DM-electron scattering process $e^-\chi\to e^-\chi$:
\begin{equation}
    i\mathcal{M}=\bar{u}_\chi(p')\left[\frac{ig_\chi}{4m_\chi^2}\gamma^\mu\gamma^5(q_\alpha q^\alpha g_{\mu\nu}-q_\nu q_\mu)\right]u_\chi(p)\frac{-ig^{\nu\rho}}{(p'-p)^2-m_{A'}^2}\bar{u}_e(k')ig_e\gamma_\rho u_e(k).
\end{equation}
Taking the non-relativistic limit we obtain the amplitude as a function of the non-relativistic operators listed in Table~\ref{tab:NRops}: 
\begin{equation}
\label{eq:scatt_1}
    i\mathcal{M}\approx 2ig_\chi g_e\frac{m_e^2}{q^2+m_{A'}^2}\frac{m_e}{m_\chi}\left[\left(1+\frac{1}{\sqrt{\beta}}\frac{m_e}{m_\chi}\right)\frac{\mu_{\chi e}}{m_e}\left(\frac{i\vec{q}}{m_e}\cdot\vperp\right)\On{11}+\frac{q^2}{m_e^2}\On8-\frac{q^2}{m_e^2}\On9\right]\, .
\end{equation}
Here $\langle\mathcal{O}_i\rangle=\langle \xi_2|\mathcal{O}_i|\xi_1\rangle$ is the amplitude of $\mathcal{O}_i$ between initial and final spin states $\xi_{1,2}$. We define the following Wilson coefficients 
\begin{align}
    &c_8=2g_\chi g_e\frac{q^2}{q^2+m_{A'}^2}\frac{m_e}{m_\chi} \nonumber\\
    &c_9=-2g_\chi g_e\frac{q^2}{q^2+m_{A'}^2}\frac{m_e}{m_\chi}\\
    &c_{11}=2g_\chi g_e\frac{m_e^2}{q^2+m_{A'}^2}\frac{\mu_{\chi e}} {m_\chi}\left(1+\frac{1}{\sqrt{\beta}}\frac{m_e}{m_\chi}\right) \nonumber
\end{align}
to obtain
\begin{equation}
    \mathcal{M}=c_8\On8+c_9\On9+c_{11}\left(\frac{i\vec{q}}{m_e}\cdot\vperp\right) \On{11}\, .
\end{equation}
We now expand the matrix element as a first order power series in $\vperp$ and find
\begin{align}
    \mathcal{M}_Sf_S &=-c_9 \left<\vec{S}_\chi\cdot\left(\frac{i\vec{q}}{m_e}\times\vec{S}_e\right)\right>f_S
\\
    \vec{\mathcal{M}}_V\cdot \vec{f}_V &=c_8\langle\vec{S}_\chi\cdot\vec{f}_V\mathbb{1}_e\rangle+c_{11}\left(\frac{i\vec{q}}{m_e}\cdot\vec{f}_V\right)\left\langle\vec{S}_\chi\cdot\frac{i\vec{q}}{m_e}\mathbb{1}_e\right\rangle\,.
\end{align}
The total amplitude to first order in $\vperp$ is given by summing the terms
\begin{equation}
    \overline{|\mathcal M_{1 \rightarrow 2} |^2}=\overline{|\mathcal{M}_Sf_S|^2}+\overline{|\mathcal{\vec{M}}_V\cdot\vec{f}_V|^2}+2\, \text{Re}\left(\overline{\vec{\mathcal{M}}_V\cdot\vec{f}_V\mathcal{M}_S^\star f_S^\star}\right)\,.
\end{equation}
Keeping only the terms that are linear in $\vperp$ yields
\begin{equation}
    \overline{|\mathcal{M}_Sf_S|^2}=\frac18|c_9|^2\frac{q^2}{m_e^2}|f_S|^2\, ,
\end{equation}
\begin{equation}
    \overline{|\mathcal{\vec{M}}_V\cdot\vec{f}_V|^2}=\frac14|c_8|^2 |\vec{f}_V|^2+\frac{\frac14|c_{11}|^2\frac{q^4}{m_e^4}}{q^2/m_e^2}\left|\frac{\vec q}{m_e}\cdot\vec{f}_V\right|^2+\frac12\text{Re}\left(c_8c_{11}^\star\right) \left|\frac{\vec q}{m_e}\cdot\vec{f}_V\right|^2\, ,
\end{equation}
and
\begin{equation}
 2\, \text{Re}\left(\overline{\vec{\mathcal{M}}_V\cdot\vec{f}_V\mathcal{M}_S^\star f_S^\star}\right)=0\, .
\end{equation}
The resulting $a_i$ coefficients are listed in Table \ref{tab:a_coeffs}.

\subsection{Majorana Dark Matter}
For our next example, we consider the case of a Majorana fermion, $\chi$, with a vector mediator, $A'^\mu$. The renormalizable interaction terms are
\begin{align}
    \mathcal{L}\supset &-\frac12 \lambda_1\bar{\chi}\gamma^\mu\gamma^5\chi A'_\mu-h_1\bar{e}\gamma^\mu e A'_\mu-h_2\bar{e}\gamma^\mu\gamma^5e A'_\mu\,.
\end{align}
We find the amplitude for DM-electron scattering in Feynman gauge to be
\begin{equation}
    i\mathcal{M}=\bar{u}_\chi(p')i(\lambda_1\gamma_\mu\gamma^5)u_\chi(p)\frac{-ig^{\mu\nu}}{(p'-p)^2-m_{A'}^2}\bar{u}_e(k')i(h_1\gamma_{\nu}+h_2\gamma_\nu\gamma^5) u_e(k)\, .
\end{equation}
In the non-relativistic limit, this reduces to 
\begin{equation}
    i\mathcal{M}\approx -i\frac{8m_\chi m_e}{q^2+m_{A'}^2}\left[\lambda_1h_1\On{8}-\lambda_1h_1\On9-2\lambda_1h_2\On4\right]\, .
\end{equation}
Next, we define the following Wilson coefficients
\begin{align}
    &c_4= \frac{16m_\chi m_e}{q^2+m_{A'}^2}\lambda_1 h_2 \nonumber\\
    &c_8= -\frac{8m_\chi m_e}{q^2+m_{A'}^2}\lambda_1 h_1\\
    &c_9= \frac{8m_\chi m_e}{q^2+m_{A'}^2}\lambda_1 h_1 \nonumber
\end{align}
to obtain
\begin{equation}
\label{eq:scatt_2}
    \mathcal{M}=c_4\On4+c_8\On8+c_9\On9\, .
\end{equation}
This yields
\begin{equation}
    \mathcal{M}_Sf_S=c_4\left\langle\vec{S}_\chi\cdot\vec{S}_e\right\rangle f_S-c_9\left<\vec{S}_\chi\cdot\left(\frac{i\vec{q}}{m_e}\times\vec{S}_e\right)\right> f_S\, ,
\end{equation}
and
\begin{equation}
    \vec{\mathcal{M}}_V\cdot\vec{f}_V=c_8\left\langle\vec{S}_\chi\cdot\vec{f}_V\mathbb{1}_e\right\rangle\, .
\end{equation}
Thus, we have
\begin{equation}
    \overline{|\mathcal{M}_Sf_S|^2}=\frac{3}{16}|c_4|^2|f_S|^2+\frac18|c_9|^2\frac{q^2}{m_e^2}|f_S|^2\, ,
\end{equation}
\begin{equation}
    \overline{|\mathcal{\vec{M}}_V\cdot\vec{f}_V|^2}=\frac14|c_8|^2|
    \vec{f}_V|^2\, ,
\end{equation}
and
\begin{equation}
    2\, \text{Re}\left(\overline{\vec{\mathcal{M}}_V\cdot\vec{f}_V\mathcal{M}_S^\star f_S^\star}\right)=0\, .
\end{equation}
The resulting $a_i$ coefficients are listed in Table \ref{tab:a_coeffs}.

\subsection{Scalar (Inelastic) Dark Matter}
Next, we study a case that populates all $a_i$ coefficients, including the imaginary part of $a_4$. In this model the DM is a complex scalar $S_1$ with mass $m_S$, but there is another dark scalar $S_2$ with mass $m_S+\delta$, with a small mass splitting $\delta \ll m_S$.
Coupling the dark scalars to the SM electron via a scalar mediator $\phi$ and a vector mediator $A^\prime_\mu$, the allowed interactions are:
\begin{align}
        \mathcal{L}\supset & - g_1^\star m_S S_1^\dagger S_2 \phi - g_1 m_S S_2^\dagger S_1\phi\nonumber \\
        &- i g_2 S_1^\dagger (\partial_\mu S_2) A^{'\mu} + i g_2^\star (\partial_\mu S_2^\dagger) S_1 A^{'\mu} -i g_3 S_2^\dagger (\partial_\mu S_1)A^{'\mu} + i g_3^\star (\partial_\mu S_1^\dagger)S_2 A^{'\mu} \\
    &-i h_2 \bar{e}\gamma^5 e \phi - h_1 \bar{e}e \phi - h_3 (\bar{e}\gamma_\mu e)A^{' \mu} - h_4 (\bar{e}\gamma_\mu\gamma^5 e)A^{'\mu}\, ,\nonumber
\end{align}
We allow the $U(1)_D$ gauge symmetry to be broken both by the mass term of the $A^\prime_\mu$ and by the couplings $g_2,g_3$ of the DM to the $A^\prime_\mu$.
In Feynman gauge, we find the scattering amplitude for $S_1e^- \to S_2 e^-$ to be
\begin{align}
    i\mathcal{M} = &(i g_1 m_S) \left ( \frac{i}{(p^\prime-p)^2-m_{\phi}^2} \right ) \bar{u}^{r^\prime}(k^\prime)i(h_1+ ih_2\gamma^5)u^r(k)\nonumber\\
    &+ (ig_3 p_\mu + i g_2^\star p_\mu^\prime)\left ( \frac{-ig^{\mu\nu}}{q^2-m^2_{A'}}\right) \bar{u}^{r^\prime}(k^\prime) (i h_3 \gamma_\nu + i h_4 \gamma_\nu \gamma^5)u^r(k)\,.
\end{align}
For inelastic DM, $\vperp$ is no longer Galilean invariant as defined in Eq. (\ref{eq:vperp}). Here we define a new Galilean invariant velocity,

\begin{equation}
    \vin = \frac{\vec{p}}{2 m_S}+\frac{\vec{p}\,^\prime}{2 (m_S+\delta)} - \frac{\vec{k} + \vec{k}^\prime}{2m_e}\, .
\end{equation}
We remark that $\vec{q}$ itself is not Galilean invariant. Since positive $\delta$ is kinematically constrained to be $\delta\sim O(v^2)$, focusing on this case allows us to neglect terms appearing at order $O(v^3)$ that break the Galilean symmetry.

In the center of momentum frame, the previously defined kinematic quantities receive the following corrections at leading order in $\epsilon=\delta/m_S$:
\begin{align}
    \vec{p} &=-\vec{k} = \mu_{Se} \vin + \frac{\vec{q}}{2}\left(1+\epsilon\frac{\mu_{Se}}{m_S}\right)\\
    \vec{p}\,^\prime &= \vec{k}' =\mu_{Se} \vin- \frac{\vec{q}}{2}\left(1-\epsilon\frac{\mu_{Se}}{m_S}\right)\\
    p^0 &= E_p \simeq m_S \sqrt{\beta} + \frac{q^2 \mu_{Se}}{4 \sqrt{\beta}m_S^2}\epsilon + \frac{1}{\sqrt{\beta}}\frac{\mu_{Se}}{2m_S} (\vec{q}\cdot \vin)\\
    p^{\prime 0} &= E_{p^\prime} \simeq m_S \sqrt{\beta} - \frac{q^2 \mu_{Se}}{4 \sqrt{\beta}m_S^2}\epsilon - \frac{1}{\sqrt{\beta}}\frac{\mu_{Se}}{2m_S} (\vec{q}\cdot \vin)\, .
\end{align}
Thus, in the non-relativistic limit, we find 
\begin{equation}
\label{eq:scatt_3}
    \mathcal{M}\simeq \left [c_1^0 + c_1^{\perp} \left ( \frac{i\vec{q}}{m_e}\cdot \vin \right ) \right ] \On1 + c_3 \On3 + c_7 \On7 + c_{10} \On{10}\, ,
\end{equation}
where the Wilson coefficients are given by
\begin{align}
\label{eq:c1_inel}
    c_1^0 &= \frac{2 m_S m_e}{q^2 + m_\phi^2}g_1 h_1 - \frac{m_S m_e}{q^2 + m_{A^\prime}^2}h_3 \left [2 (g_3 + g_2^\star)\sqrt{\beta} + (g_3-g_2^\star)\epsilon\frac{q^2\mu_{Se}}{2m_S^2m_e}\left(1+\frac{m_e}{\sqrt{\beta}m_S}\right) \right ] \\
    c_1^{\perp} &= \frac{im_Sm_e}{q^2 + m_{A^\prime}^2}h_3 (g_3 - g_2^\star) \frac{\mu_{Se}}{m_S} \left (\frac{m_e}{m_S\sqrt{\beta}} + 1 \right ) \\
    c_3 &= \frac{2 m_S m_e}{q^2 + m^2_{A^\prime}}h_3 (g_3 + g_2^\star)\frac{\mu_{Se}}{m_S} \\
    c_7 &= \frac{4 m_S m_e}{q^2 + m_{A^\prime}^2}h_4  (g_3 + g_2^\star)\frac{\mu_{Se}}{m_S} \left ( \frac{m_S}{m_e}\sqrt{\beta} + 1\right )  \\
    \label{eq:c10_inel}
    c_{10} &= \frac{-2m_S m_e}{q^2 + m_\phi^2} g_1 h_2 - 2ih_4\frac{ m_S m_e}{q^2 + m_{A^\prime}^2}\left[(g_3+g_2^\star)\epsilon \frac{\mu_{Se}}{m_S} \frac{m_e}{m_S}\left ( 1 + \frac{m_S}{m_e}\sqrt{\beta} \right ) +(g_3 - g_2^\star) \frac{m_e}{m_S}\right]\, .
\end{align}
This yields
\begin{align}
    \mathcal{M}_S f_S &\simeq \left ( c_1^0 \left\langle \mathbb{1}_S \mathbb{1}_e\right\rangle + c_{10} \Big \langle \mathbb{1}_S \frac{i\vec{q}}{m_e}\cdot \vec{S}_e \Big\rangle\right )\, , 
\end{align}
\begin{align}
    \vec{\mathcal{M}}_V \cdot \vec{f}_V &\simeq c_1^{\perp}\left(\vec{q}\cdot \vec{f}_V\right) \langle \mathbb{1}_S\mathbb{1}_e\rangle + c_3 \Big \langle \mathbb{1}_S \left ( \frac{i\vec{q}}{m_e} \times \vec{f}_V \right )\cdot \vec{S}_e \Big \rangle + c_7 \left\langle \mathbb{1}_S \vec{f}_V \cdot \vec{S}_e\right\rangle\, ,
\end{align}
such that
\begin{align}
\overline{|\mathcal{M}_S f_S|^2} &= \left ( |c_1^0|^2 + \frac{|c_{10}|^2}{4} \frac{\vec{q}^2}{m_e^2} \right ) |f_S|^2 \, ,
\end{align}
\begin{align}
\overline{|\vec{\mathcal{M}}_V\cdot\vec{f}_V|^2} =& \left ( |c_1^{\perp}|^2 - \frac{|c_3|^2}{4} \right )\Bigg | \frac{\vec{q}}{m_e}\cdot \vec{f}_V \Bigg |^2 + \frac{1}{4} \left ( |c_3|^2\frac{\vec{q}^2}{m_e^2} + |c_7|^2 \right ) |\vec{f}_V|^2 \nonumber \\
&+ i\frac{1}{2} \text{Re}(c_3 c_7^\star) \frac{\vec{q}}{m_e}\cdot \left(\vec{f}_V \times \vec{f}_V^\star\right)\, ,
\end{align}
and
\begin{equation}
2\, \text{Re} \overline{\left ( \mathcal{M}_S f_S (\mathcal{M}_V\cdot \vec{f}_V)^\star\right )} = 2\, \text{Im} \left [ \left ( c_1^0 c_1^{\perp\star} - \frac14 c_{10}c_7^\star \right ) f_S \left ( \frac{\vec{q}}{m_e}\cdot \vec{f}_V^\star\right ) \right ]\, .
\end{equation}
In general, $a_4$ is complex. In particular, the phase in $a_4$ is not simply generated from $g_2,g_3$ being complex since even if $g_2$ and $g_3$ were both real, a relative phase would still be present in $c_{10}$.
In the case of elastic scattering, $\epsilon \to 0$, and since we have only one species of DM, $g_3 \to g_2$, so $a_4$ becomes real.
For $a_4 \neq 0$, (i) both scalar and vector mediators should be present, or (ii) if only the vector mediator is present, at least one of $g_2,g_3$ should be complex or DM should be inelastic, as can be seen in Eqs.~(\ref{eq:c1_inel}--\ref{eq:c10_inel}). The different possibilities are summarized in Table \ref{tab:inelastic_a4_coeff}.
\begin{table}[h]
    \centering
    \resizebox{\textwidth}{!}{
    \begin{tabular}{|c||c|c|c|c|c |} \hline
         Model & $a_0$ & $a_1$ & $a_2$ & $a_3$ & $a_4$  \\\hline 
         \small{Anapole} & $\frac{1}{8}|c_9|^2 x_e$ & $\frac{1}{4}|c_8|^2  $ & $\frac{1}{4}|c_{11}|^2 x_e^2+\frac{1}{2}{\rm Re}(c_8 c_{11}^\star)$ & 0 & 0  \Tstrut\Bstrut \\ 
         \small{Majorana} & $\frac{3}{16}|c_{4}|^2 +\frac{1}{8}|c_9|^2x_e$ & $\frac{1}{4}|c_8|^2$ & 0 & 0 & 0 \Tstrut\Bstrut\\ 
         \small{Scalar Inelastic} & $|c_{1}^0|^2 +\frac{1}{4}|c_{10}|^2 x_e$ & $\frac{1}{4}|c_{3}|^2 x_e+\frac{1}{4}|c_7|^2$ & ${\color{red}|{c_1^\perp}|^2}x_e-\frac{1}{4}|c_{3}|^2 x_e$ & $\frac{1}{2}{\rm Re}(c_3 c_{7}^\star)$ & $c_1^0{\color{red}c_1^{\perp\star}}-\frac{1}{4}c_{10} c_7^\star$ \Tstrut\Bstrut \\ \hline 
    \end{tabular}
    }
    \caption{The values of $a_i$, for each of the three case studies in Section~\ref{sec:casestudies}. 
    For conciseness, we define $x_e \equiv (q/m_e)^2$. The coefficients $c_1,\ldots,c_{15}$ are those corresponding to the effective non-relativistic operators $\mathcal{O}_1,\ldots,\mathcal{O}_{15}$ defined in Table \ref{tab:NRops}, as they appear in the scattering amplitudes \eqref{eq:scatt_1}, \eqref{eq:scatt_2}, and \eqref{eq:scatt_3}. We note that in the case of scalar elastic DM, the $c_1^\perp$ terms highlighted in {\color{red}red} will disappear and all coefficients $c_i$ are real.}
    \label{tab:a_coeffs}
\end{table}

\begin{table}[h!]
    \centering
    \resizebox{\textwidth}{!}{
    \begin{tabular}{|c|c|c|c|c|c|c|}
    \hline
    {\bf Scalar Inelastic DM} & ~ & $c_1^0$ & $c_1^\perp$ & $c_7$ & $c_{10}$ & $a_4$ \Tstrut\Bstrut \\ \hline
    \multirow{3}{*}{\begin{tabular}{c} (i) Scalar and vector \\ mediators present \end{tabular}} & $g_2,g_3$ real & real & imaginary & real & complex & complex \Tstrut\Bstrut \\ \cline{2-7}
     & elastic ($\epsilon\to 0$, $g_3\to g_2$) & real & real & real & real & real \Tstrut\Bstrut \\ \cline{2-7}
     & $g_2,g_3$ real and elastic & real & zero & real & real & real \Tstrut\Bstrut \\ \hline
    \multirow{3}{*}{\begin{tabular}{c}(ii) Only vector mediator \\ present   ($h_1=h_2=0$) \end{tabular}} & $g_2,g_3$ real & real & imaginary & real & imaginary & imaginary \Tstrut\Bstrut \\ \cline{2-7}
     & elastic ($\epsilon\to 0$, $g_3\to g_2$) & real & real & real & real & real \Tstrut\Bstrut \\ \cline{2-7}
     & $g_2,g_3$ real and elastic & real & zero & real & zero & zero \Tstrut\Bstrut \\ \hline
    \end{tabular}
    }
    \caption{Starting from the generic scalar inelastic DM model, we show how the real and/or imaginary parts of $a_4$ can vanish in certain limits. If $g_2,g_3$ are complex and $\epsilon \neq 0$, all four coefficients $c_1^0, c_1^\perp, c_7, c_{10}$ are complex, leading to complex $a_4$ in both cases (i) and (ii). Note that if the vector mediator is removed, then $a_4$ is identically zero in this model.}
    \label{tab:inelastic_a4_coeff}
\end{table}

Finally, we emphasize that both the electron vector and axial vector currents must couple to the spin-1 mediators in order to generate $a_3$.

\section{Discussion and Conclusions}

In this work, we have extended the general framework for computing spin-dependent DM scattering rates beyond the conventional assumption of isotropic detector materials, establishing a consistent set of conventions applicable to anisotropic targets. We find that the most general squared amplitude governing nonrelativistic DM–SM interactions can be expressed in terms of the seven distinct material response functions and five DM response functions that appear in \eqref{eq:a01234B1234567}. 
Despite the fact that these material response functions depend entirely on SM physics, it is often computationally challenging to calculate them from first principles. 
Under common simplifying conditions, namely parity conservation, isotropy, or electronic adiabaticity, we have uncovered new relations between these functions, reducing the computational burden of the spin-dependent scattering rate prediction.

While a reduction in computational effort is always welcome, 
our simplified form of the spin-dependent scattering cross section  
from \eqref{eq:aEtcB0} also provides valuable insights for the design of future detectors. For example, a parity-symmetric detector medium is completely insensitive to the $a_3$ DM response function. A pair of direct detection experiments with chiral and achiral detector targets would be able to isolate the $a_3$ component of a DM signal.
Similarly, as we demonstrated in \eqref{eq:a013}, the $a_2$ and $a_4$ contributions to the rate both vanish in electronically adiabatic systems: to detect DM scattering through these DM response functions, a detector medium must have substantial difference between the Coulomb potentials in its initial and final states.

Altogether, our results provide a systematic foundation for incorporating anisotropic material effects into DM direct detection calculations, and a road map for connecting microscopic DM models to experimentally accessible observables.

\section*{Acknowledgments}
We thank Carlos Blanco and M. Grant Roberts for helpful discussion. We also thank Carlos Blanco and Tongyan Lin for useful comments on a draft of this manuscript. The research of P.G. and P.M. is supported in part by the U.S. Department of Energy grant number DE-SC0010107. The research of P.M. is supported in part by the UC Chancellor's Dissertation Year Fellowship. The work of B.L. was supported in part by the U.S. Department of Energy under Grant Number DE-SC0011640. T-TY is supported in part by NSF CAREER grant PHY-1944826. T-TY thanks the hospitality of the Università degli Studi di Padova and the CERN Theory group where portions of this work were completed. 

\appendix
\section{Explicit computation of form factors} \label{app:explicit}
In this appendix, we verify explicitly that the electronic adiabaticity relation in \eqref{eq:qdotA},  $\vec{q}\cdot\vec{f}_{1\to 2} = \left [ \Delta E - q^2/(2m_e) \right ]f_S$ with $\Delta E = E_2 - E_1$, holds for a particle-in-a-box model and the hydrogen atom.

\subsection{Particle in a box}
For a rectangular box of dimensions $(L_x, L_y, L_z)$ and volume $V = L_x L_y L_z$, the eigenstates are given by
\begin{equation}
    \psi_{\vec{n}} = \frac{2^{3/2}}{\sqrt{V}} \sin \frac{\pi n_x x}{L_x} \sin \frac{\pi n_y y}{L_y} \sin \frac{\pi n_z z}{L_z}
\end{equation}
where $\vec{n} = (n_x, n_y, n_z)$, with $n_i\in \mathbb{N}$, labels the possible excited states  starting $(1,1,1)$ as the ground state. If $\psi_1$ is the ground state and $\psi_2$ is a state with arbitrary $\vec{n}$, then the excited state energy is given by
\begin{equation}
    \Delta E = \frac{\pi^2}{2 m_e} \left ( \frac{n_x^2 - 1}{L_x^2} + \frac{n_y^2 - 1}{L_y^2} + \frac{n_z^2 - 1}{L_z^2}\right )\, .
\end{equation}
Since the integral for the scalar form factor factorizes, we can write
\begin{equation}
    f_S = f_x(q_x) f_y (q_y) f_z(q_z)
\end{equation}
where 
\begin{equation}
    f_j(q_j) = \int_0^L dx_j \, e^{iq_j x_j} \frac{2}{L_j} \sin\left (\frac{\pi n_j x_j}{L_j}\right ) \sin \left ( \frac{\pi x_j}{L_j} \right )
\end{equation}
for $j=x,y,z$. We compute each $f_j(q_j)$ and find
\begin{equation}
    f_j(q_j) = -i4\pi^2 n_j q_j L_j \frac{[1+(-1)^{n_j}e^{i q_j L_j}]}{[(q_j L_j)^2 - \pi^2(n_j -1)^2][(q_j L_j)^2 - \pi^2 (n_j + 1)^2]}\, .
\end{equation}
Similarly, the vector form factor factorizes
\begin{equation}
    \vec{f}_{1\to 2} = f_{\nabla}^{(x)}(q_x)f_y(q_y) f_z(q_z)\, \hat{x} + f_x(q_x) f_{\nabla}^{(y)}(q_y) f_z(q_z)\,\hat{y} + f_x(q_x) f_y(q_y) f_{\nabla}^{(z)}(q_z) \, \hat{z}\, ,
\end{equation}
where 
\begin{equation}
     f_{\nabla}^{(j)}(q_j) = -\frac{i}{m_e} \int_0^L dx_j\, e^{i q_j x_j} \frac{2}{L_j} \frac{\pi}{L_j} \sin \left ( \frac{\pi n_j x_j}{L_j} \right ) \cos \left ( \frac{\pi x_j}{L_j} \right )\, ,
\end{equation}
resulting in
\begin{equation}
    f_{\nabla}^{(j)}(q_j) = i \frac{2\pi^2 n_j}{m_e L_j} \frac{[1 + (-1)^{n_j}e^{i q_j L_j}][(q_j L_j)^2 - \pi^2 (n_j^2-1)]}{[(q_j L_j)^2 - \pi^2 (n_j-1)^2][(q_j L_j)^2 - \pi^2(n_j+1)^2]}\, .
\end{equation}
Noting that
\begin{equation}
    \frac{f_\nabla^{(j)}(q_j)}{f_j(q_j)} = \frac{1}{2 m_e q_j L_j^2} [  \pi^2(n_j^2-1)-(q_j L_j)^2]
\end{equation}
it is then straightforward to verify the relation $\vec{q}\cdot \vec{f}_{1\to 2} = [\Delta E - q^2/(2m_e)]f_S$ holds.
\subsection{Hydrogen atom}
Here we consider the $1/r$ potential of the hydrogen atom.
Using the formulae in the appendices of~\cite{DarkSide:2018ppu,Catena:2019gfa}, the form factors for the hydrogen atom can be found for a transition from an initial state $|n,l,m\rangle$ to a final state $|n^\prime, l^\prime, m^\prime\rangle$. In order for this section to be self-contained, we recall here the expressions for the form factors:
\begin{equation}
    f_S = \sqrt{4\pi} \sum_{L = |l-l^\prime|}^{l+l^\prime} i^L I_1^L \sum_{M= - L}^L Y_L^{M\star} (-1)^{m^\prime} \sqrt{(2l+1)(2l^\prime+1)(2L+1)} \begin{pmatrix} l & l^\prime & L \\ 0 & 0 & 0\end{pmatrix} \begin{pmatrix} l & l^\prime & L \\ m & - m^\prime & M \end{pmatrix}
\end{equation}
\begin{multline}
\label{eq:vec_form_hyd}
    \vec{f}_{1\to 2} = -i \frac{\sqrt{4\pi}}{m_e}\sum_{i=1}^3 \hat{e}_i \sum_{\hat{l}=l-1}^{l+1} \sum_{\hat{m}=m-1}^{m+1} \sum_{L=0}^\infty \sum_{M=-L}^L Y_L^{M\star} i^L (-1)^{m^\prime} \sqrt{(2\hat{l}+1)(2l^\prime + 1)(2L+1)} \\ \begin{pmatrix} \hat{l} & l^\prime & L \\ 0 & 0 & 0\end{pmatrix} \begin{pmatrix} \hat{l} & l^\prime & L \\ \hat{m} & - m^\prime & M \end{pmatrix} (c_{\hat{l}\hat{m}}^{(i)} I_2^L + d_{\hat{l}\hat{m}}^{(i)}I_3^L)
\end{multline}
where $Y_L^M$ are the spherical harmonics and $c_{\hat{l}\hat{m}}^{(i)}$ and $d_{\hat{l}\hat{m}}^{(i)}$ are both constants that only depend on the summation indices.\footnote{Note that there is a relative minus sign difference between the formula in Ref.~\cite{Catena:2019gfa} and the definition in \eqref{eq:Fconvert}.}
The energy difference between the states is given by
\begin{equation}
    \Delta E = -\frac{1}{2m_e a^2} \left (\frac{1}{n'^2} - \frac{1}{n^2} \right )
\end{equation}
where $m_e$ is the electron mass and $a$ is the Bohr radius.

For a central potential, the radial integrals are factorized. Ref.~\cite{Catena:2019gfa} gives the following expressions
\begin{align}
    I_1^L &= \int_0^\infty dr\, r^2 R^\star_{n^\prime l^\prime}R_{nl} j_L(qr)  \\
    I_2^L &= \int_0^\infty dr\, r^2 R^\star_{n^\prime l^\prime} \frac{dR_{nl}}{dr} j_L(qr) \\
    I_3^L &= \int_0^\infty dr \, r R^\star_{n^\prime l^\prime} R_{nl} j_L(qr) 
\end{align}
where $L$ is the angular momentum (e.g. for the scalar form factor $L = |l-l^\prime|, \ldots,l+l^\prime$). Due to the form of hydrogen radial wavefunctions, we can write
\begin{align}
    I_1^L &= \sum_{k=l+l^\prime}^{(n-1)+(n^\prime-1)}\alpha_k^{nl,n^\prime l^\prime} \int_0^\infty dr\, r^2 \left (\frac{r}{a}\right)^k e^{-r/(pa)} j_L(qr)\\
    I_2^L &= \sum_{\substack{k=l+l^\prime-1 \\ k\geq 0}}^{(n-1)+(n^\prime-1)}\beta_k^{nl,n^\prime l^\prime} \int_0^\infty dr\, r^2 \left (\frac{r}{a}\right)^k e^{-r/(pa)} j_L(qr)\\
    I_3^L &= \sum_{\substack{k=l+l^\prime-1 \\ }}^{(n-1)+(n^\prime-1)-1}\gamma_k^{nl,n^\prime l^\prime} \int_0^\infty dr\, r^2 \left (\frac{r}{a}\right)^k e^{-r/(pa)} j_L(qr)\, .
\end{align}
with $p = (1/n + 1/n^\prime)^{-1}$. Here, $\alpha_k, \beta_k, \gamma_k$ are coefficients that depend only on $n,l,n^\prime,l^\prime$, and $a$. We will suppress the superscripts $nl,n^\prime l^\prime$ henceforth for ease of notation. $I_1^L$ is the radial integral for the scalar form factor $f_S$ whereas the vector form factor has contributions from both $I_2^L$ and $I_3^L$. 
The building block for these integrals is analytically computed to be
\begin{equation}
    \int_0^\infty dr\, r^2 \left (\frac{r}{a}\right )^k e^{-r/(pa)} j_L(qr) = h^L_k \, {}_2F_{1} \left ( \frac{L-k}{2},\frac{L-k-1}{2}, \frac{3}{2}+L, -X \right )
\end{equation}
with ${}_2 F_1$ the hypergeometric function, $X=a^2p^2q^2$, and where we define the coefficient $h^L_k$ to be
\begin{equation}
    h^L_k \equiv 2^{-1-L} p^{3+L+k} a^{3+L} \sqrt{\pi} \frac{q^L}{(1+X)^{2+k}} \frac{\Gamma(3+L+k)}{\Gamma(3/2+L)}\, .
\end{equation}

In the following, $\phi$ is the azimuthal angle of $\vec{q}$ and we employ the notation $c_\theta \equiv \cos\theta$ and $s_\theta \equiv \sin \theta$ where $\theta$ is the polar angle of the vector $\vec{q}$.
As an example, taking $n^\prime=3,\, l^\prime = 2,\, m^\prime = 2$ and $n = 2,\, l=1,\, m=1$, the radial wavefunctions are 
\begin{align}
    R_{21} &= \mathcal{N}_{21} a^{-3/2}\frac{r}{a}e^{-r/(2a)} \\
    R_{32} &= \mathcal{N}_{32} a^{-3/2} \left (\frac{r}{a}\right )^2 e^{-r/(3a)}
\end{align}
with $\mathcal{N}_{21} = 1/(2\sqrt{6})$ and $\mathcal{N}_{32} = 4/(81\sqrt{30})$.
The non-zero coefficients are given by
\begin{align}
    \alpha_3 &= \mathcal{N}_{21}\mathcal{N}_{32} a^{-3} \\
    \beta_2 &= \alpha_3/a\\
    \beta_3 &= -\frac{1}{2}\alpha_3/a\\
    \gamma_2 &= \alpha_3/a \, .
\end{align}
We find for the scalar form factor
\begin{multline}
\label{eq:scalar_form_hyd}
    f_S = \sqrt{4\pi} \alpha_3 \sum_{\substack{L=1 \\}}^3 h_3^L i^L \sqrt{15 (2L+1)} \begin{pmatrix} 1 & 2 & L \\ 0 & 0 & 0  \end{pmatrix} \begin{pmatrix} 1 & 2 & L \\ 1 & -2 & 1 \end{pmatrix} Y_L^{1 \star} \, {}_2F_1 \left ( \frac{L-3}{2}, \frac{L-4}{2}, \frac{3}{2} + L, -X \right )
\end{multline}
where $p=6/5$. Here the $L=2$ term will not contribute to the summation as the Wigner $3-j$ symbol evaluates to 0.
We simplify each of the components of the vector form factor in turn. We obtain for the $x,y$ components
\begin{align}
\label{eq:vec_form_hyd_xy}
    &f_{1\to2,(x,y)} = i\frac{\sqrt{4\pi}}{m_e} (Y_2^2)^\star h^{2}_2 \left [(c_{00}^{(x,y)} \beta_2 +d_{00}^{(x,y)}\gamma_2) + c_{00}^{(x,y)} \beta_3 \frac{p}{(1+X)} (7-X) \right ] \\
    &- \left[\frac{i\sqrt{4\pi}}{m_e} \sum_{\hat{m}=0,2}~\sum_{L=0,2,4} g^L(\hat{m})h^{L}_2 (Y_L^{2-\hat{m}})^\star c_{2\hat{m}}^{(x,y)}\beta_3 \frac{p}{(1+X)}(5+L) \, {_2 F_{1}} \left ( \frac{L-3}{2}, \frac{L-4}{2}, \frac{3}{2}+L, -X\right )\right]
\end{align}
where the first term comes from the $\hat{l}=0, L=2, M=2$ term and the rest is from the $\hat{l}=2$ portion of ~\eqref{eq:vec_form_hyd}. Similarly, the $z$-component is given by
\begin{multline}
\label{eq:vec_form_hyd_z}
    f_{1\to 2,z} = -i\frac{\sqrt{4\pi}}{m_e} \sum_{L= 2,4}\, (Y_L^1)^\star g^L(1) h^L_2 c_{21}^{(z)}\beta_3 \frac{p}{(1+X)}(5+L) \, {}_2 F_{1} \left ( \frac{L-3}{2}, \frac{L-4}{2}, \frac{3}{2}+L, -X\right )
\end{multline}
where we have also used that $ c_{2\hat{m}}^{(x,y)}\beta_2 + d_{2\hat{m}}^{(x,y)}\gamma_2 = c_{21}^{(z)}\beta_2 + d_{21}^{(z)}\gamma_2 = 0 $ for $\hat{m}=0,2$.
Evaluating the Wigner $3j$-symbols for these values of $L,\hat{m}$ gives 
\begin{align}
    g^L(\hat{m}) &= (-1)^{\hat{m}}2\sqrt{6} (L^3+3L^2-10L-24) \frac{(L-1)}{(4-L)!(5+L)!} \sqrt{\frac{(2+L-\hat{m})!(2+\hat{m})!}{(2-\hat{m})!(L-2+\hat{m})!}} i^L 5\sqrt{(2L+1)} 
\end{align}
while the $c^{(i)}_{\hat{l}\hat{m}},d^{(i)}_{\hat{l}\hat{m}}$ coefficients are defined as in Ref.~\cite{Catena:2019gfa}:
\begin{align}
c_{00}^{(x)} &= -ic_{00}^{(y)} = -\frac{1}{\sqrt{6}} \\
d_{00}^{(x)} &= -i d_{00}^{(y)} = - \sqrt{\frac{2}{3}} \\
c_{20}^{(x)} &= - i c_{20}^{(y)} = \frac{1}{\sqrt{30}}\\
c_{22}^{(x)} &= i c_{22}^{(y)} = -\frac{1}{\sqrt{5}} \\ 
c_{21}^{(z)} &= \frac{1}{\sqrt{5}}\, .
\end{align}
For $L\leq 4$, the hypergeometric function evaluates to a polynomial. This allows us to make the replacement
\begin{align}
    {}_2 F_{1} \left ( \frac{L-3}{2}, \frac{L-4}{2}, \frac{3}{2}+L, -X\right ) &\to 1 - X \frac{(L-4)(L-3) [20+8L -X(L-1)(L-2)]}{8(3+2L)(5+2L)}
\end{align}
in~\eqref{eq:scalar_form_hyd}, \eqref{eq:vec_form_hyd_xy} and~\eqref{eq:vec_form_hyd_z}.

Numerically, the scalar form factor evaluates to
\begin{equation}
    f_S = -\frac{829440 i a q e^{-i \phi } s_\theta \left(36 a^2 q^2 c_{2 \theta}+25\right)}{\left(36 a^2 q^2+25\right)^5}
\end{equation}
whereas the results in Cartesian coordinates for the vector form factor are given by 
\begin{equation}
    f_{1\to 2,x} = -\frac{2304 i \left[\left(25-18 a^2 q^2\right)^2+36a^2 q^2 e^{-2 i \phi } \left(3 e^{2 i \phi } \left(9 a^2 q^2 c_{4 \theta} +25 c_{2 \theta}\right)+36 a^2 q^2 s^2_{\theta}(1-3 c_{2 \theta})-50 s^2_{\theta }\right)\right]}{a m_e \left(36a^2 q^2+25\right)^5}
\end{equation}

\begin{equation}
f_{1\to 2,y} = - i f_{1\to 2,x} - 331776a q^2 e^{-2 i \phi} \frac{[25 s^2_\theta - 18 a^2 q^2 s^2_\theta (1-3c_{2\theta})]}{ m_e \left(36a^2 q^2+25\right)^5}
\end{equation}
\begin{equation}
    f_{1\to2,z} = \frac{248832 i a q^2 e^{-i \phi } s_{2 \theta} \left(36 a^2 q^2 c_{2 \theta}+25\right)}{m_e \left(36 a^2 q^2+25\right)^5}\, .
\end{equation}
This gives
\begin{align}
\vec{q}\cdot \vec{f}_{1\to 2} & =  11520 i qs_\theta e^{-i \phi} \frac{(36 a^2 q^2 - 5) (36 a^2 q^2 c_{2\theta} + 25)}{a m_e (36 a^2 q^2 + 25)^5}\\
    & = \left(\frac{5}{72a^2m_e}-\frac{q^2}{2m_e}\right)\left(-\frac{829440 i a q e^{-i \phi } s_\theta \left(36 a^2 q^2 c_{2 \theta}+25\right)}{\left(36 a^2 q^2+25\right)^5}\right)\\
    & = \left [ \Delta E - \frac{q^2}{2m_e} \right ]f_S\, ,
\end{align}
satisfying \eqref{eq:qdotA}. Similarly, we can verify that the relation holds for other initial and final states.

\paragraph{Parity:} 
We also verify the statement made in Sec~\ref{sec:parity} namely that if $\psi_{1,2}$ have matching parity, then $f_S$ is purely real whereas if $\psi_{1,2}$ have opposite parities, then $f_S$ is purely imaginary. 

As an example, we will take $|\psi_2\rangle = |3,\, 2,\,1\rangle + |3,\, 2,\, -1\rangle$ (parity even) and $|\psi_1\rangle = |2,\, 1,\, 1\rangle + | 2,\, 1,\, -1\rangle$ (parity odd). This gives
\begin{equation}
    f_S = i\frac{1658880 a q c_\theta \left[72 a^2 q^2 \left(2 s^2_\theta c_{2\phi} + c_{2 \theta}\right )-36 a^2 q^2+25\right]}{\left(36 a^2 q^2+25\right)^5}
\end{equation}
which is purely imaginary.

On the contrary, if we take $|\psi_2\rangle = |3,\, 1,\, 1\rangle + |3,\, 1,\, -1\rangle$ (parity odd) keeping $\psi_1$ the same, we obtain
\begin{equation}
    f_S = \frac{9216[-54 a^2 q^2 (36 a^2 q^2 -175)(2s^2_\theta c_{2\phi} + c_{2\theta})+648 a^4q^4 -5850 a^2 q^2 + 3125]}{\sqrt{5}(36 a^2 q^2 + 25)^5}
\end{equation}
which is purely real. We also note that both of the above examples satisfy the relation of \eqref{eq:parity_q_opp}.

\bibliography{whim_refs}

\end{document}